\documentclass[%
aps,
reprint,
preprintnumbers,
nofootinbib,
 amsmath,amssymb,
]{revtex4-2}

\usepackage{graphicx}
\usepackage{dcolumn}
\usepackage{bm}
\usepackage{slashed} 
\usepackage{hyperref}
\hypersetup{pdfstartview=FitV,colorlinks=true,linkcolor=blue,citecolor=red,filecolor=black,urlcolor=blue}
\usepackage[dvipsnames]{xcolor}
\usepackage{subfig}
\usepackage{ulem}

\newcommand{\lrpartial}{\overset{\leftrightarrow}{\partial}}

\begin{document}

\preprint{OU--HET--1206}

\title{Gravitational Waves from Particle Decays during Reheating}

\author{Shinya Kanemura}
    \email{kanemu@het.phys.sci.osaka-u.ac.jp}
    \affiliation{Department of Physics, Osaka University, Toyonaka, Osaka 560-0043, Japan}
\author{Kunio Kaneta}
    \email{kaneta@het.phys.sci.osaka-u.ac.jp}
    \affiliation{Department of Physics, Osaka University, Toyonaka, Osaka 560-0043, Japan}

\date{\today}

\begin{abstract}
    Gravitational waves have become an irreplaceable tool for exploring the post-inflationary universe.
    Their cosmological and astrophysical origins have been attracting numerous attention.
    In this Letter, we point out a novel source of ultra-high frequency gravitational waves: the decay of particles produced during the reheating era.
    We highlight the decay of the Higgs boson as 
    a representative case, showing how it yields a testable gravitational wave spectrum by future observations.
\end{abstract}

\maketitle

%\tableofcontents

\section{Introduction}
Inflation stands as a cornerstone in modern cosmology~\cite{Olive:1989nu,Linde:1990flp,Lyth:1998xn,Linde:2000kn,Martin:2013tda,Martin:2013nzq,Martin:2015dha} as it has been supported by various observations, such as big bang nucleosynthesis (BBN)~\cite{Alpher:1948ve,Walker:1991ap,Olive:1999ij} and cosmic microwave background (CMB)~\cite{Akrami2020}.
Yet, there is still a missing piece between the inflationary epoch and the radiation-dominated universe.
The missing link, known as reheating, is a crucial process through which new phenomena beyond the Standard Model (SM), such as the dark matter production~\cite{Hall:2009bx,Chu:2011be,Mambrini:2013iaa,Chu:2013jja,Kaneta:2016wvf,Kaneta:2017wfh,Bernal:2017kxu,Biswas:2018aib,Kaneta:2019zgw,Bernal:2019mhf,Bernal:2020gzm,Bernal:2020qyu,Anastasopoulos:2020gbu,Brax:2020gqg,Brax:2021gpe,Kaneta:2021pyx,Ghosh:2022hen} and leptogenesis~\cite{Giudice:1999fb,Asaka:1999yd,Kaneta:2019yjn}, may take place.

Reheating also sources high-frequency cosmological gravitational waves (GWs).
For instance, the scattering~\cite{Ema:2020ggo,Klose:2022knn} and the decay~\cite{Nakayama2019,Barman2023,Barman:2023rpg} of the inflaton during reheating in general produce
GWs, where the inflaton is a scalar field causing inflation.
When the inflationary scale is sufficiently high, 
the generated GWs can be imprinted at high-frequency bands.
Low-frequency GWs are, on the other hand, often predicted in scenarios including the first-order phase transition~\cite{Apreda:2001us,Grojean:2006bp,Espinosa:2008kw,Caprini:2015zlo,Kakizaki:2015wua,Hashino:2016rvx,Hashino:2016xoj,Athron:2023xlk} and the formation of the topological defects~\cite{Vilenkin:1981bx,Vachaspati:1984gt,Caldwell:1991jj,Damour:2000wa,Damour:2001bk,Hiramatsu:2013qaa}.

In this Letter, we explore a novel source of GWs that may 
provide insights into the dynamics of a nonthermal particle during reheating.
As testable scenarios, we consider decays of inflaton and the Higgs boson in the SM and its extension after inflation, resulting in GWs at frequencies higher than those aimed by LISA~\cite{LISA:2017pwj}, DECIGO~\cite{Seto:2001qf}, BBO~\cite{Harry:2006fi}, but accessible at other proposed experiments searching for high-frequency GWs~\cite{Aggarwal:2020olq,Herman:2020wao,Berlin:2021txa,Domcke:2022rgu,Herman2022,Bringmann:2023gba}.

The Letter will proceed as follows.
After clarifying the inflationary sector in Sec.~\ref{sec: Inflationary sector} and evaluating the GW spectrum produced from the inflaton decay based on a Starobinsky-type model~\cite{Starobinsky1980} as an example in Sec.~\ref{sec: GW spectrum from inflaton decay}, we apply a similar method to the case of the decay of the Higgs boson produced by the inflaton decay during reheating in the SM in Sec.~\ref{sec: GW spectrum from the SM Higgs boson decay} and its extension in Sec.~\ref{sec: GW spectrum from the B-L Higgs boson decay}.
Finally, we summarize the results in Sec.~\ref{sec: Results} and give discussions and conclusions in Sec.~\ref{sec: Discussions and conclusions}.

\section{Inflationary sector}
\label{sec: Inflationary sector}
As a viable example of inflation, we consider a Starobinsky-type model~\cite{Starobinsky1980}, whose potential for the inflaton $\phi$ is given by
\begin{align}
    V(\phi) &= \frac{3}{4}m_\phi^2M_P^2\left(
        1-e^{-\sqrt{\frac{2}{3}}\frac{\phi}{M_P}}
        \right)^2,
\end{align}
where $M_P\simeq 2.4\times10^{18}$ GeV is the reduced Planck mass, and $m_\phi$ is the inflaton mass, which is determined by the amplitude of the curvature power spectrum $A_S$ as
\begin{align}
    m_\phi^2 &\simeq \frac{24\pi^2 A_S}{N_*^2}M_P^2,
\end{align}
where $N_*$ is the number of e-folds when the pivot scale $k_*=0.05~{\rm Mpc}^{-1}$ exits the horizon during inflation.
From the Planck data~\cite{Akrami2020}, the amplitude of the scalar perturbation is given by $A_S=2.099\times 10^{-9}$.
By taking $N_*=50$ for definiteness, we obtain $m_\phi\simeq 3\times 10^{13}$ GeV.

After inflation is over at the scale factor $a=a_e$, the inflaton starts oscillating about the potential minimum where approximately $V(\phi)\simeq m_\phi^2\phi^2/2$.
The inflaton energy density, $\rho_\phi$, follows the Boltzmann equation,
\begin{align}
    \dot\rho_\phi + 3H\rho_\phi = -\Gamma^{\rm (E)}_\phi\rho_\phi,
    \label{eq: Boltzmann eq for rho_phi}
\end{align}
where $H=\dot a/a$ is the Hubble parameter, and the energy transfer rate $\Gamma^{\rm (E)}_\phi$ is defined by~\cite{Garcia2020,Garcia2020a}
\begin{align}
    \Gamma^{\rm (E)}_\phi &= \frac{1}{\rho_\phi} \sum_{n=1}^\infty \int E_n|{\cal M}_n|^2 d{\rm LIPS},
\end{align}
with $n$ being a label of the $n$-th Fourier mode of the inflaton oscillation.
The decay amplitude of the $n$-th mode is denoted by ${\cal M}_n$ with the transferred energy $E_n=nm_\phi$.
The Lorentz-invariant phase-space element $d{\rm LIPS}$ should be taken according to the final state kinematics.

We suppose that the inflaton predominantly decays into a pair of the SM fermions ($f$), whose interaction is given by
\begin{align}
    {\cal L}_{\phi,{\rm int}} &= y_\phi \phi \overline f f.
\end{align}
We note that the energy transfer rate in the present case coincides with the normal decay width of the inflaton,
\begin{align}
    \Gamma_{\phi\to\overline f f} &= \frac{y_\phi^2}{8\pi}m_\phi.
\end{align}
The reheating completes when $\rho_\phi(a_{\rm reh})=\rho_{\rm rad}(a_{\rm reh})$, where $\rho_{\rm rad}$ represents the radiation energy density.
The reheating temperature is estimated as~\cite{Garcia2020}
\begin{align}
    T_{\rm reh}
    &\simeq
    7.7\times10^4~{\rm GeV}\times \left(
        \frac{427/4}{g_{\rm reh}}
    \right)^{1/4}
    \left(
        \frac{y_\phi}{10^{-10}}
    \right),
\end{align}
where $g_{\rm reh}=g_*(T_{\rm reh})=427/4$ if all the SM particles are in the thermal bath at $T=T_{\rm reh}$.

\section{GW spectrum from inflaton decay}
\label{sec: GW spectrum from inflaton decay}
\begin{figure}[t]
    \centering
    \includegraphics[width=.2\textwidth]{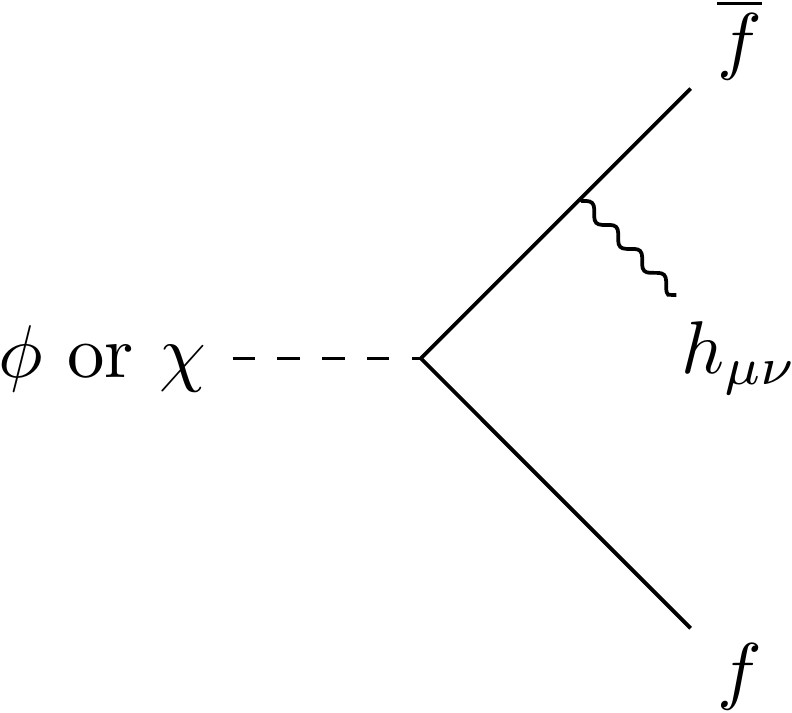}~~~
    \includegraphics[width=.2\textwidth]{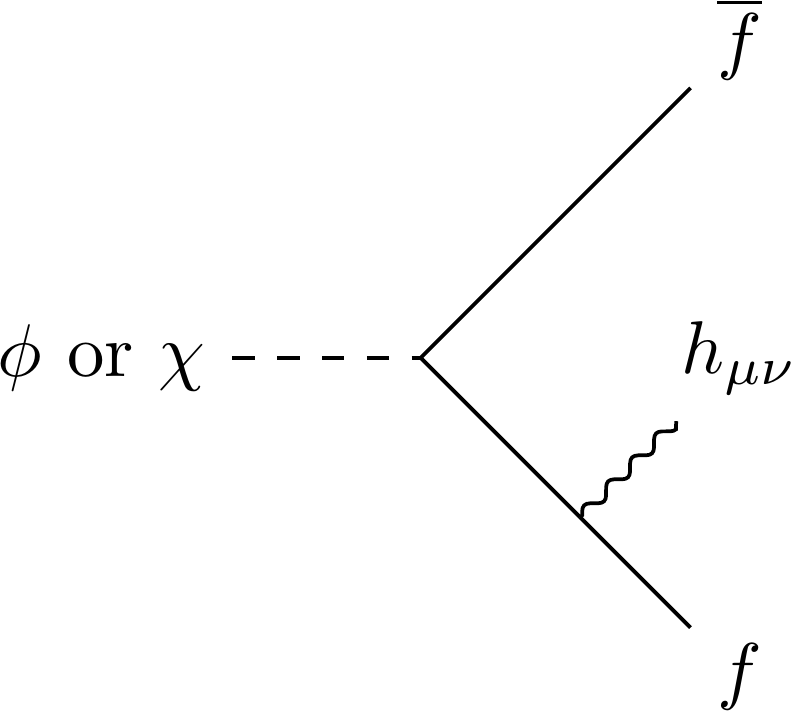}
    \caption{Graviton bremsstrahlung with the initial state being either the inflaton or the Higgs boson.}
    \label{fig: diagrams}
\end{figure}

Associated with the decay into fermions, a single graviton is emitted via the bremsstrahlung process~\cite{Nakayama2019,Barman2023,Barman:2023rpg}.
The relevant diagrams for the three-body decay are shown in Fig.~\ref{fig: diagrams}.
The total decay width can then be divided into two pieces,
\begin{align}
    \Gamma_\phi &= \Gamma_\phi^{(0)} + \Gamma_\phi^{(1)},
\end{align}
where $\Gamma^{(0)}_{\phi}\simeq\Gamma_{\phi\to\overline f f}$ does not involve gravitons in the final state, while $\Gamma^{(1)}_\phi$ has a single graviton in addition to the fermions.
It is convenient to further decompose $\Gamma^{(1)}_\phi$ into two pieces~\cite{Barman2023},
\begin{align}
    \Gamma^{(1)}_\phi &= \Gamma^{(1)}_{\phi\to {\rm rad}} + \Gamma^{(1)}_{\phi\to {\rm GW}},\\
    \Gamma^{(1)}_{\phi\to {\rm rad}} &= \frac{1}{\rho_\phi}\sum_{n=1}^\infty\int (E_n-E_{\rm GW})|{\cal M}^{(1)}_n|^2d{\rm LIPS},\\
    \Gamma^{(1)}_{\phi\to {\rm GW}} &= \frac{1}{\rho_\phi}\sum_{n=1}^\infty\int E_{\rm GW}|{\cal M}^{(1)}_n|^2d{\rm LIPS},
\end{align}
with ${\cal M}^{(1)}_n$ being the corresponding amplitude, where $E_{\rm GW}$ is the energy of graviton.

The evolution of the graviton energy density, $\rho_{\rm GW}$, follows the Boltzmann equation,
\begin{align}
    \dot\rho_{\rm GW} + 4H\rho_{\rm GW} &= \Gamma^{(1)}_{\phi\to {\rm GW}}\rho_\phi,
\end{align}
from which we obtain
\begin{align}
    a_{\rm reh}^4\frac{d\rho_{\rm GW}(a_{\rm reh})}{d E_{\rm GW}} &= \int_{a_e}^{a_{\rm reh}}da \frac{a^3}{H}\left[\frac{d\Gamma^{(1)}_{\phi\to {\rm GW}}}{dE_{\rm GW}}\rho_\phi\right].
\end{align}
We note that $d\Gamma^{(1)}_{\phi\to{\rm GW}}/dE_{\rm GW}$ is defined as the differential decay width of the inflaton,
\begin{align}
    \frac{d\Gamma^{(1)}_{\phi\to {\rm GW}}}{dE_{\rm GW}}
    &\equiv
    \frac{E_{\rm GW}}{\rho_\phi}\sum_{n=1}^\infty\int |{\cal M}^{(1)}_n|^2\frac{d{\rm LIPS}_3}{dE_{\rm GW}},
\end{align}
where $d{\rm LIPS}_3/dE_{\rm GW}$ indicates the integration for the three-body phase space, without integrating for $E_{\rm GW}$.
The gravitational wave spectrum can be obtained by
\begin{align}
    h^2\Omega_{\rm GW} &=
    \frac{1}{\rho_{\rm cr,0}h^{-2}}\frac{d\rho_{\rm GW}(a_{\rm reh})}{d\ln E_{\rm GW}}\left(
        \frac{a_{\rm reh}}{a_0}
    \right)^4,
    \label{eq: h2OmegaGW}
\end{align}
with $\rho_{\rm cr,0}h^{-2}=8.0992\times10^{-47}~{\rm GeV}^4$ being the present critical density.

Based on the above argument, we can readily evaluate the GW spectrum associated with the inflaton decay into fermions.
The graviton $h_{\mu\nu}$ is defined by $g_{\mu\nu}= \eta_{\mu\nu}+2h_{\mu\nu}/M_P$ with $\eta_{\mu\nu}={\rm diag}(+1,-1,-1,-1)$, and couples to the energy-momentum tensor as
\begin{align}
    {\cal L}_{h_{\mu\nu},{\rm int}} &=
    -\frac{h_{\mu\nu}}{M_P}T^{\mu\nu}.
\end{align}
For a spin-1/2 particle we have
\begin{align}
    T^{1/2}_{\mu\nu} &= \frac{i}{4}\overline f(\gamma_\mu\lrpartial_\nu+\gamma_\nu\lrpartial_\mu)f,
\end{align}
where $\overline f\lrpartial_\mu f\equiv \overline f\partial_\mu f - (\partial_\mu \overline f)f$.
We have here neglected the terms proportional to $\eta_{\mu\nu}$ as it vanishes due to the traceless condition of the graviton.
After summing over all the spin and the polarization states,
we obtain
the differential energy-transfer rate as
\begin{align}
    \frac{d\Gamma^{(1)}_{\phi\to{\rm GW}}}{dE_{\rm GW}}
    &=
    \frac{y_\phi^2m_\phi^2}{64\pi^3M_P^2}F(E_{\rm GW}/m_\phi),
\end{align}
where $F(x)=(1-2x)(1-2x+2x^2)$ under the approximation of $m_f\ll m_\phi$.
Thus, we arrive at
\begin{align}
    \Omega_{\rm GW} 
    &\simeq 
    \frac{y_\phi^2m_\phi^2\rho_e E_{\rm GW}}{160\pi^3H_e M_P^2\rho_{\rm cr,0}}
    \left(
        \frac{a_{\rm reh}}{a_e}
    \right)^{-3/2}
    \left(
        \frac{a_0}{a_{\rm reh}}
    \right)^{-4},\label{eq: OmegaGW from inflaton decay}
\end{align}
for $E_{\rm GW}\ll m_\phi$, where $\rho_\phi(a_e)\equiv\rho_e\simeq 0.175m_\phi^2M_P^2$ in the Starobinsky model~\cite{Ellis2015}, and $H(a_e)\equiv H_e=\sqrt{\rho_e/3}/M_P\simeq 7.2\times10^{12}~{\rm GeV}\times(\rho_e/0.175m_\phi^2M_P^2)^{1/2}$.
Using $E_{\rm GW} = 2\pi f (a_0/a_{\rm reh})$ with $f$ and $a_0$ being the frequency of the GWs in the present universe and the today's scale factor, respectively, we obtain
\begin{align}
    h^2\Omega_{\rm GW} &\simeq
    1.4\times10^{-36}(f/{\rm Hz})\left(\frac{y_{\phi}}{10^{-10}}\right),
    \label{eq: h2OGW from inflaton decay}
\end{align}
with the peak frequency being
\begin{align}
    f_{\rm peak} &=
    \frac{m_\phi}{4\pi}\left(\frac{a_{\rm reh}}{a_0}\right)
    \simeq
    3.7\times10^{18}\left(
        \frac{y_\phi}{10^{-10}}
    \right)^{-1}~{\rm Hz}.
    \label{eq: f_peak from inflaton decay}
\end{align}
Our estimation is consistent with the results in Ref.~\cite{Barman2023}.

\section{GW spectrum from the SM Higgs boson decay}
\label{sec: GW spectrum from the SM Higgs boson decay}
We now turn to a discussion of the GWs generated by the Higgs boson decay after the end of inflation.
We suppose that the Higgs doublet $\Phi$ is produced from the inflaton decay through the interaction term,
\begin{align}
    {\cal L}_{\phi,{\rm int}} &= \mu \phi \Phi^\dagger\Phi,
    \label{eq: inflaton-Higgs coupling}
\end{align}
where $\mu$ is a dimensionful coupling.
By denoting $\chi$ as a real scalar degree of freedom, the decay rate of $\phi\to \chi\chi$ is given by
\begin{align}
    \Gamma_{\phi\to \chi\chi} &= \frac{\mu^2}{32\pi m_\phi}.
\end{align}
Thus, the branching fraction is obtained as
\begin{align}
    {\rm Br}_\chi &\simeq \frac{4\Gamma_{\phi\to \chi\chi}}{\Gamma_{\phi\to\overline f f}}
    =
    y_\phi^{-2}\frac{\mu^2}{m_\phi^2},
\end{align}
where a factor of 4 counts the number of real scalar degrees of freedom in $\Phi$.
In the following, we take ${\rm Br}_\chi$ as a free parameter.
We note that the coupling (\ref{eq: inflaton-Higgs coupling}) induces a temporal contribution to the squared mass of the Higgs boson given by $\mu\sqrt{2\rho_\phi/m_\phi^2}$. In the parameter space of our interest, the temporal mass contribution is sufficiently smaller than $m_\phi$, otherwise, the nonperturbative effect becomes important~\cite{Garcia2020a}.

The produced Higgs boson decays into gauge bosons and fermions.
To illustrate how the Higgs boson decay generates GWs, for the sake of brevity, we assume that the Higgs boson predominantly decays into a pair of fermions through the Yukawa interactions.
By writing the relevant Yukawa coupling as $y_f$, the decay width of $\chi$ into a pair of fermions is given by
\begin{align}
    \Gamma_{\chi\to\overline f f} &= \frac{y_f^2}{16\pi}m_\chi,
\end{align}
where $m_\chi$ is the Higgs boson mass during reheating.
Both $m_\chi$ and $y_f$ are to be renormalized values at the relevant scale of the decay at early times, while we take them as free parameters based on the agnostic viewpoint, as explained below.

For simplicity, we consider the case where the squared Higgs boson mass is positive so that the electroweak (EW) symmetry is not broken spontaneously during reheating; namely, the masses of the four real scalar degrees of freedom in $\Phi$ are $m_\chi$.
The following argument, however, does not depend on whether the Higgs field acquires a vacuum expectation value or not, as long as the nonthermal condition is satisfied, as described in the following paragraph.
Although the Higgs boson mass today is $m_\chi\simeq 125$ GeV, it can be much larger during reheating because of the unbroken EW symmetry.

When the Higgs mass during reheating is in excess of the temperature, the produced Higgs boson maintains a nonthermal spectrum \cite{Garcia:2018wtq,Ballesteros:2020adh,Garcia:2021iag}, which is given by
\begin{align}
    f_\chi(p,t) &\simeq 
    \frac{24\pi^2n_\chi(t)}{m_\phi^3}
    \left(
        \frac{m_\phi}{2p}
    \right)^{3/2}\theta(m_\phi/2-p),
\end{align}
where $p$ denotes the absolute value of the Higgs momentum, and $\theta(x)$ is the Heaviside step function.
The number density of the Higgs doublet, $n_\chi(t)$, follows the Boltzmann equation given by 
\begin{align}
    \dot n_\chi + 3H n_\chi = {\rm Br}_\chi \Gamma_\phi n_\phi(t),
    \label{eq: Boltzmann eq for n_h}
\end{align}
where the inflaton number density is obtained by solving Eq.~(\ref{eq: Boltzmann eq for rho_phi}), yielding $n_\phi=(\rho_e/m_\phi)(a/a_e)^{-3}$ for $a\ll a_{\rm reh}$.
Integrating Eq.~(\ref{eq: Boltzmann eq for n_h}) from $a_e$ to $a$, we obtain
\begin{align}
    n_\chi(a) &= \frac{2{\rm Br}_\chi\Gamma_\phi\rho_e}{3m_\phi H_e}\left(\frac{a}{a_e}\right)^{-3}\left[
        \left(
            \frac{a}{a_e}
        \right)^{3/2}-1
    \right].
\end{align}
Note that when the decay rate of $\chi$ becomes non-negligible, $n_\chi$ starts decreasing as $a^{-3}$, resulting in a suppression in the GW production.
The time scale of when this happens can be estimated from ${\rm Br}_\chi\Gamma_\phi n_\phi(a)\simeq \Gamma_\chi (m_\chi/m_\phi) n_\chi(a)$\footnote{The dilation factor  $m_\chi/m_\phi$ is included. See Eq.~(\ref{eq: gamma}) for comparison.} at $a=a_d$, yielding
\begin{align}
    \frac{a_d}{a_e}
    &\simeq
    \left(
    \frac{2\Gamma_\chi m_\chi}{3H_e m_\phi}
    \right)^{-2/3}.
\end{align}
In the following, we require $a_d>a_{\rm reh}$, corresponding to $\Gamma_\phi m_\phi\gtrsim \Gamma_\chi m_\chi$.

In a similar manner to the inflaton decay, a single graviton is emitted via the bremsstrahlung process in the Higgs boson decay, as shown in Fig.~\ref{fig: diagrams}.
By taking the nonthermal Higgs distribution into account, the Boltzmann equation for $\rho_{\rm GW}$ is written as
\begin{align}
    a^{-3}H\frac{d}{da}\left[
        a^4 \frac{d\rho_{\rm GW}}{dE_{\rm GW}}
    \right]
    &=
    \frac{d\Gamma^{(1)}_{\chi\to{\rm GW}}}{dE_{\rm GW}} \gamma,\label{eq: Boltzmann eq for rho_GW from Higgs}
\end{align}
where $d\Gamma^{(1)}_{\chi\to{\rm GW}}/dE_{\rm GW}$ is the differential decay rate of the Higgs boson at rest, and
\begin{align}
    \gamma &\equiv E_{\rm GW}
    \int \frac{d^3p}{(2\pi)^3} \frac{m_\chi}{p^0}f_\chi(p,t)
    \simeq 6\frac{m_\chi}{m_\phi}E_{\rm GW}n_\chi(t),
    \label{eq: gamma}
\end{align}
for $m_\chi\ll m_\phi$.
The differential decay rate is given by
\begin{align}
    \frac{d\Gamma^{(1)}_{\chi\to{\rm GW}}}{dE_{\rm GW}} &=
    \frac{y_f^2m_\chi^3}{128\pi^3M_P^2E_{\rm GW}}F(E_{\rm GW}/m_\chi),
\end{align}
where the kinetic factor $F(x)$ is the same as the inflaton decay width.
Integrating Eq.~(\ref{eq: Boltzmann eq for rho_GW from Higgs}) from $a=a_e$ to $a=a_{\rm reh}$, we obtain
\begin{align}
    \frac{d\rho_{\rm GW}(a_{\rm reh})}{dE_{\rm GW}}
    &=
    \frac{y_f^2{\rm Br}_\chi\Gamma_\phi m_\chi^4 \rho_e}{640\pi^3H_e^2m_\phi^2M_P^2}F(E_{\rm GW}/m_\chi)\left(
        \frac{a_{\rm reh}}{a_e}
    \right)^{-4}\nonumber\\
    &\times
    \left[
        5\left(\frac{a_{\rm reh}}{a_e}\right)^4-8\left(\frac{a_{\rm reh}}{a_e}\right)^{5/2}+3
    \right].
    \label{eq: drho_GW/dEG from Higgs decay}
\end{align}
For $a_{\rm reh}\gg a_e$ and $E_{\rm GW}\ll m_\chi$, we arrive at
\begin{align}
    &
    h^2\Omega_{\rm GW}
    \simeq 
    \frac{1}{\rho_{\rm cr,0}h^{-2}}
    \frac{y_f^2{\rm Br}_\chi\Gamma_\phi m_\chi^4 \rho_e E_{\rm GW}}{128\pi^3H_e^2m_\phi^2M_P^2}
    \left(
        \frac{a_0}{a_{\rm reh}}
    \right)^{-4}\nonumber\\
    \simeq&
    5.5\times 10^{-22}\left(\frac{f}{\rm Hz}\right)
    y_f^2{\rm Br}_\chi\left(\frac{m_\chi}{10^{12}~{\rm GeV}}\right)^4\left(
        \frac{y_\phi}{10^{-10}}
    \right)^{-1},
    \label{eq: h2OGW from Higgs decay}
\end{align}
with the peak frequency given by
\begin{align}
    f_{\rm peak} 
    &\simeq
    1.2\times10^{17}\left(
        \frac{m_\chi}{10^{12}~{\rm GeV}}
    \right)\left(
        \frac{y_\phi}{10^{-10}}
    \right)^{-1}~{\rm Hz}.
    \label{eq: f_peak from Higgs decay}
\end{align}

It is important to notice that in Eq.~(\ref{eq: drho_GW/dEG from Higgs decay}), $d\rho_{\rm GW}(a_{\rm reh})/dE_{\rm GW}$ is independent from $a_{\rm reh}/a_e$
at leading order for $a_{\rm reh}\gg a_e$, while the same quantity in the case of the inflaton decay becomes $d\rho_{\rm GW}(a_{\rm reh})/dE_{\rm GW}\propto(a_{\rm reh}/a_e)^{-3/2}$ as can be seen in Eq.~(\ref{eq: OmegaGW from inflaton decay}); i.e., a larger suppression than the case of the Higgs boson decay.
The difference comes from the fact that, in the case of the Higgs boson decay, the inflaton 
keeps producing 
the Higgs boson during reheating, enhancing the GW production from the Higgs boson decay, and hence the dilution effect for $\rho_{\rm GW}$ is mitigated.
Such an enhancement is absent in the case of inflaton decay as the inflaton is the primary source of producing GWs.

On the other hand, by noticing that the maximal GW amplitude is obtained when $a_d\simeq a_{\rm reh}$, corresponding to $\Gamma_\phi m_\phi\simeq \Gamma_\chi m_\chi$.
In this case, we obtain $h^2\Omega_{\rm GW}(f_{\rm peak})\simeq 1.2\times10^{-19}{\rm Br}_\chi$.
This result implies that the ratio of the energy densities of radiation and GW is crucial, suggesting the following new venue for new physics to be tested by GWs.

\section{GW spectrum from the B-L Higgs boson decay}
\label{sec: GW spectrum from the B-L Higgs boson decay}
As an example of new physics models, we consider the gauged $B-L$ model where U(1)$_{B-L}$ is gauged by adding three right-handed neutrinos.
An additional Higgs field $S$, charged +2 under U(1)$_{B-L}$, is introduced to spontaneously break U(1)$_{B-L}$ by acquiring a non-zero VEV, $\langle S\rangle = v_S$.
In the broken phase where $S=v_S+s/\sqrt{2}$, the real part $s$ obtains a mass $m_s$, while the right-handed neutrinos and the U(1)$_{B-L}$ gauge boson acquire masses $m_N$ and $m_{Z'}$, respectively.
For simplicity, we focus on the case where the spectrum of these new particles are $m_s\lesssim m_\phi \ll m_{Z'},m_N$ and where the inflaton decays only into a pair of $s$.

By integrating out $N$ and $Z'$, $s$ obtains an effective coupling to the Riemann tensor \cite{Ema:2021fdz},
\begin{align}
    {\cal L}_{s,2h} 
    &=
    \beta_{\alpha_3}\frac{s}{v_S}R^{\mu\nu\rho\sigma}R_{\mu\nu\rho\sigma},
\end{align}
with $\beta_{\alpha_3}=31/11520\pi^2$.
Thus, the decay width of $s$ into a pair of gravitons is
\begin{align}
    \Gamma_{s\to{\rm GW}}
    &=
    \frac{\beta_{\alpha_3}}{4\pi}\frac{m_s^7}{v_S^2M_P^4}.
\end{align}
The differential decay width becomes
\begin{align}
    \frac{d\Gamma_{s\to{\rm GW}}}{dE_{\rm GW}}
    &=
    \Gamma_{s\to{\rm GW}}
    \left(
    \frac{2E_{\rm GW}}{m_s}
    \right)^2
    \delta(E_{\rm GW}-m_s/2).
\end{align}
On the other hand, $s$ may also decay into four SM fermions through off-shell $Z'$, $s\to Z'^* Z'^*\to f\bar f f' \bar f'$, which is the leading decay channel for the given mass spectrum.
Taking the massless limit for the final state fermions, we obtain the decay width as
\begin{align}
    \Gamma_{s\to{\rm rad}}
    &\simeq
    \frac{529}{11796480\pi^4}\frac{m_s^7}{v_S^6},
    \label{eq: Gamma s -> rad}
\end{align}
where we have used $m_{Z'}\gg m_s$.

In a similar manner to the SM Higgs case, we estimate the GW spectrum as
\begin{align}
    \Omega_{\rm GW}
    &\simeq
    \frac{5}{2}\frac{\Gamma_{s\to{\rm GW}}}{\Gamma_{s\to{\rm rad}}}
    \left(
    \frac{4\pi f}{m_s}
    \right)^{5/2}
    \left(
    \frac{a_0}{a_{\rm reh}}
    \right)^{5/2}
    \left(
    \frac{a_{\rm eq}}{a_0}
    \right),
\end{align}
where $a_{\rm eq}$ is the matter-radiation equality time, namely, $a_{\rm eq}/a_0 \simeq 1/3388$~\cite{Planck:2018vyg}.
At the peak frequency
\begin{align}
    f_{\rm peak}
    &=
    \frac{m_s}{4\pi}\frac{a_{\rm reh}}{a_0}\nonumber\\
    &\simeq 
    7.3\times10^{26}~{\rm Hz}
    \left(
    \frac{m_s}{10^{13}~{\rm GeV}}
    \right)^{-5/2}
    \left(
    \frac{v_S}{M_P}
    \right)^3,
\end{align}
we obtain
\begin{align}
    \Omega_{\rm GW}(f_{\rm peak})
    &=
    \frac{5}{2}\frac{\Gamma_{s\to{\rm GW}}}{\Gamma_{s\to{\rm rad}}}\left(
    \frac{a_{\rm eq}}{a_0}
    \right)
    \simeq
    9.5\times10^{-6}\left(
    \frac{v_S}{M_P}
    \right)^4.
\end{align}

\begin{figure}[t]
    \centering
    \includegraphics[width=.5\textwidth]{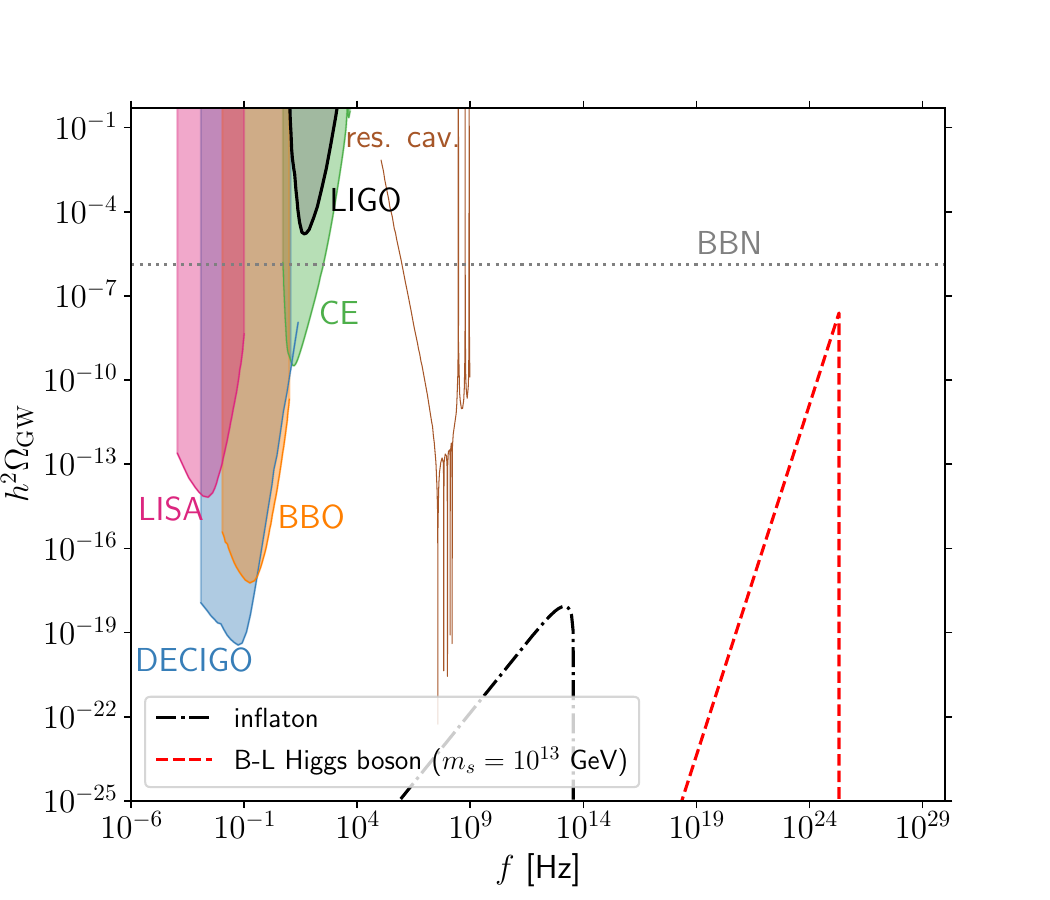}
    \caption{Summary of the GW spectrum from the decays of the inflaton and the $B-L$ Higgs boson.}
    \label{fig: summary}
\end{figure}

\section{Results}
\label{sec: Results}
In Fig.~\ref{fig: summary} we show our results, where we have taken $y_\phi=10^{-5}$ for the inflaton decay (black dot-dashed) and $v_S=0.3M_P$ for the $B-L$ Higgs decay (red dashed).
As can be seen from Eqs.~(\ref{eq: h2OGW from inflaton decay}) and (\ref{eq: f_peak from inflaton decay}), the peak value of $h^2\Omega_{\rm GW}\simeq 5.2\times10^{-18}$ from the inflaton decay does not depend on $y_\phi$, while the place of the peak varies with $y_\phi$.
The black shaded region is excluded by the LIGO observing run 3~\cite{KAGRA:2013rdx}, and the other shaded regions are future projections of LISA and BBO, whose sensitivity curves are taken from Ref.~\cite{Thrane:2013oya}.
The sensitivity curves of CE and DECIGO are taken from Refs.~\cite{Reitze:2019iox} and~\cite{Seto:2001qf}, respectively.
Shown in the brown line indicates the parameter space that can be explored by proposed experiments using resonant cavities~\cite{Herman:2020wao,Herman2022}.
The constraint from the BBN, $h^2\Omega_{\rm GW}<1.3\times10^{-6}$ \cite{Yeh:2022heq}, is depicted by the dotted line.

For the GWs from the inflaton decay, 
if we require perturbativity as $y_\phi\lesssim 10$, the lowest value of $f_{\rm peak}\simeq 3.7\times10^{7}$ Hz is suggested, which is beyond the reach of the future interferometers. The resonant cavities, on the other hand, would have a chance to test such a scenario.
The GWs from the $B-L$ Higgs boson decay can reach the BBN bound when $m_s$ and $v_S$ are sufficiently large.
On the other hand, for BBN to happen, we need $T_{\rm reh}>4~{\rm MeV}$~\cite{Hasegawa:2019jsa}, or equivalently $\Gamma_{s\to{\rm rad}}\gtrsim 10^{-23}~{\rm GeV}(T_{\rm reh}/4~{\rm MeV})^2$.
From Eq.~(\ref{eq: Gamma s -> rad}), we obtain $(m_s/10^{13}~{\rm GeV})^7(0.3M_P/v_S)^6\gtrsim(T_{\rm reh}/4~{\rm MeV})^2$, by which the peak value of $\Omega_{\rm GW}$ is limited as $h^2\Omega_{\rm GW}(f_{\rm peak})/7.9\times10^{-8}\lesssim (m_s/10^{13}~{\rm GeV})^{14/3}(T_{\rm reh}/4~{\rm MeV})^{-4/3}$.
Note that for future projections including COrE and Euclid, we expect that $h^2\Omega_{\rm GW}\gtrsim 7.6\times10^{-8}$ can be tested~\cite{Pagano:2015hma}.

\section{Discussions and conclusions}
\label{sec: Discussions and conclusions}
We briefly mention possible models that fit our argument.
As we have seen, to maximize the GWs from the Higgs decay, $m_\chi$ needs to be large enough, i.e., a large hierarchy between the Higgs boson mass during reheating and the EW scale.
This type of hierarchy requires fine-tuning in a context of, for instance, the radiative breaking of EW symmetry in supersymmetric models~\cite{Ibanez1982,Inoue1982,Inoue1982a,AlvarezGaume1983}.
One of such models is discussed in, e.g., Ref.~\cite{Ellis2018a}, where the Higgs boson mass at the EW scale is obtained by the radiative EW symmetry breaking together with cancellations among mass parameters of the order of the grand unification scale.

Finally, we emphasize that the GW spectrum discussed in this Letter is not limited to the decay of the Higgs boson.
As one such example, we have considered the $B-L$ breaking Higgs decay, in which the produced GWs may be tested in future observations.
It is also applicable to heavy particles produced during reheating, including those that emerged from supersymmetry and extended Higgs sectors.
The resultant GWs are notably characterized by ultra-high frequencies, which have recently gained increased attention, spurring numerous discussions and proposals for forthcoming experiments (see, e.g., Refs.~\cite{Carney:2023nzz,Ito:2023nkq}).
Taking particle decay during reheating as a new source of GWs will open up avenues for extensive future research.

\section*{Acknowledgments}
We would like to thank Yann Mambrini, Motohiko Yoshimura, and Liliana Velasco-Sevilla for the useful discussion.
The work of S. K. and K. K. was supported in part by JSPS KAKENHI Nos. 20H00160 (S. K. and K. K.) and 23K17691 (S. K.).

\bibliography{biblio}

%apsrev4-2.bst 2019-01-14 (MD) hand-edited version of apsrev4-1.bst
%Control: key (0)
%Control: author (8) initials jnrlst
%Control: editor formatted (1) identically to author
%Control: production of article title (0) allowed
%Control: page (0) single
%Control: year (1) truncated
%Control: production of eprint (0) enabled
\begin{thebibliography}{81}%
\makeatletter
\providecommand \@ifxundefined [1]{%
 \@ifx{#1\undefined}
}%
\providecommand \@ifnum [1]{%
 \ifnum #1\expandafter \@firstoftwo
 \else \expandafter \@secondoftwo
 \fi
}%
\providecommand \@ifx [1]{%
 \ifx #1\expandafter \@firstoftwo
 \else \expandafter \@secondoftwo
 \fi
}%
\providecommand \natexlab [1]{#1}%
\providecommand \enquote  [1]{``#1''}%
\providecommand \bibnamefont  [1]{#1}%
\providecommand \bibfnamefont [1]{#1}%
\providecommand \citenamefont [1]{#1}%
\providecommand \href@noop [0]{\@secondoftwo}%
\providecommand \href [0]{\begingroup \@sanitize@url \@href}%
\providecommand \@href[1]{\@@startlink{#1}\@@href}%
\providecommand \@@href[1]{\endgroup#1\@@endlink}%
\providecommand \@sanitize@url [0]{\catcode `\\12\catcode `\$12\catcode `\&12\catcode `\#12\catcode `\^12\catcode `\_12\catcode `\%12\relax}%
\providecommand \@@startlink[1]{}%
\providecommand \@@endlink[0]{}%
\providecommand \url  [0]{\begingroup\@sanitize@url \@url }%
\providecommand \@url [1]{\endgroup\@href {#1}{\urlprefix }}%
\providecommand \urlprefix  [0]{URL }%
\providecommand \Eprint [0]{\href }%
\providecommand \doibase [0]{https://doi.org/}%
\providecommand \selectlanguage [0]{\@gobble}%
\providecommand \bibinfo  [0]{\@secondoftwo}%
\providecommand \bibfield  [0]{\@secondoftwo}%
\providecommand \translation [1]{[#1]}%
\providecommand \BibitemOpen [0]{}%
\providecommand \bibitemStop [0]{}%
\providecommand \bibitemNoStop [0]{.\EOS\space}%
\providecommand \EOS [0]{\spacefactor3000\relax}%
\providecommand \BibitemShut  [1]{\csname bibitem#1\endcsname}%
\let\auto@bib@innerbib\@empty
%</preamble>
\bibitem [{\citenamefont {Olive}(1990)}]{Olive:1989nu}%
  \BibitemOpen
  \bibfield  {author} {\bibinfo {author} {\bibfnamefont {K.~A.}\ \bibnamefont {Olive}},\ }\bibfield  {title} {\bibinfo {title} {{Inflation}},\ }\href {https://doi.org/10.1016/0370-1573(90)90144-Q} {\bibfield  {journal} {\bibinfo  {journal} {Phys. Rept.}\ }\textbf {\bibinfo {volume} {190}},\ \bibinfo {pages} {307} (\bibinfo {year} {1990})}\BibitemShut {NoStop}%
\bibitem [{\citenamefont {Linde}(1990)}]{Linde:1990flp}%
  \BibitemOpen
  \bibfield  {author} {\bibinfo {author} {\bibfnamefont {A.~D.}\ \bibnamefont {Linde}},\ }\href@noop {} {\emph {\bibinfo {title} {{Particle physics and inflationary cosmology}}}},\ \bibinfo {series} {Contemp. Concepts Phys.}, Vol.~\bibinfo {volume} {5}\ (\bibinfo  {publisher} {Harwood},\ \bibinfo {year} {1990})\ pp.\ \bibinfo {pages} {1--362},\ \Eprint {https://arxiv.org/abs/hep-th/0503203} {hep-th/0503203} \BibitemShut {NoStop}%
\bibitem [{\citenamefont {Lyth}\ and\ \citenamefont {Riotto}(1999)}]{Lyth:1998xn}%
  \BibitemOpen
  \bibfield  {author} {\bibinfo {author} {\bibfnamefont {D.~H.}\ \bibnamefont {Lyth}}\ and\ \bibinfo {author} {\bibfnamefont {A.}~\bibnamefont {Riotto}},\ }\bibfield  {title} {\bibinfo {title} {{Particle physics models of inflation and the cosmological density perturbation}},\ }\href {https://doi.org/10.1016/S0370-1573(98)00128-8} {\bibfield  {journal} {\bibinfo  {journal} {Phys. Rept.}\ }\textbf {\bibinfo {volume} {314}},\ \bibinfo {pages} {1} (\bibinfo {year} {1999})},\ \Eprint {https://arxiv.org/abs/hep-ph/9807278} {arXiv:hep-ph/9807278} \BibitemShut {NoStop}%
\bibitem [{\citenamefont {Linde}(2000)}]{Linde:2000kn}%
  \BibitemOpen
  \bibfield  {author} {\bibinfo {author} {\bibfnamefont {A.~D.}\ \bibnamefont {Linde}},\ }\bibfield  {title} {\bibinfo {title} {{Inflationary cosmology}},\ }\href {https://doi.org/10.1016/S0370-1573(00)00038-7} {\bibfield  {journal} {\bibinfo  {journal} {Phys. Rept.}\ }\textbf {\bibinfo {volume} {333}},\ \bibinfo {pages} {575} (\bibinfo {year} {2000})}\BibitemShut {NoStop}%
\bibitem [{\citenamefont {Martin}\ \emph {et~al.}(2014{\natexlab{a}})\citenamefont {Martin}, \citenamefont {Ringeval},\ and\ \citenamefont {Vennin}}]{Martin:2013tda}%
  \BibitemOpen
  \bibfield  {author} {\bibinfo {author} {\bibfnamefont {J.}~\bibnamefont {Martin}}, \bibinfo {author} {\bibfnamefont {C.}~\bibnamefont {Ringeval}},\ and\ \bibinfo {author} {\bibfnamefont {V.}~\bibnamefont {Vennin}},\ }\bibfield  {title} {\bibinfo {title} {{Encyclop\ae{}dia Inflationaris}},\ }\href {https://doi.org/10.1016/j.dark.2014.01.003} {\bibfield  {journal} {\bibinfo  {journal} {Phys. Dark Univ.}\ }\textbf {\bibinfo {volume} {5-6}},\ \bibinfo {pages} {75} (\bibinfo {year} {2014}{\natexlab{a}})},\ \Eprint {https://arxiv.org/abs/1303.3787} {arXiv:1303.3787 [astro-ph.CO]} \BibitemShut {NoStop}%
\bibitem [{\citenamefont {Martin}\ \emph {et~al.}(2014{\natexlab{b}})\citenamefont {Martin}, \citenamefont {Ringeval}, \citenamefont {Trotta},\ and\ \citenamefont {Vennin}}]{Martin:2013nzq}%
  \BibitemOpen
  \bibfield  {author} {\bibinfo {author} {\bibfnamefont {J.}~\bibnamefont {Martin}}, \bibinfo {author} {\bibfnamefont {C.}~\bibnamefont {Ringeval}}, \bibinfo {author} {\bibfnamefont {R.}~\bibnamefont {Trotta}},\ and\ \bibinfo {author} {\bibfnamefont {V.}~\bibnamefont {Vennin}},\ }\bibfield  {title} {\bibinfo {title} {{The Best Inflationary Models After Planck}},\ }\href {https://doi.org/10.1088/1475-7516/2014/03/039} {\bibfield  {journal} {\bibinfo  {journal} {JCAP}\ }\textbf {\bibinfo {volume} {03}},\ \bibinfo {pages} {039}},\ \Eprint {https://arxiv.org/abs/1312.3529} {arXiv:1312.3529 [astro-ph.CO]} \BibitemShut {NoStop}%
\bibitem [{\citenamefont {Martin}(2016)}]{Martin:2015dha}%
  \BibitemOpen
  \bibfield  {author} {\bibinfo {author} {\bibfnamefont {J.}~\bibnamefont {Martin}},\ }\bibfield  {title} {\bibinfo {title} {{The Observational Status of Cosmic Inflation after Planck}},\ }\href {https://doi.org/10.1007/978-3-319-44769-8_2} {\bibfield  {journal} {\bibinfo  {journal} {Astrophys. Space Sci. Proc.}\ }\textbf {\bibinfo {volume} {45}},\ \bibinfo {pages} {41} (\bibinfo {year} {2016})},\ \Eprint {https://arxiv.org/abs/1502.05733} {arXiv:1502.05733 [astro-ph.CO]} \BibitemShut {NoStop}%
\bibitem [{\citenamefont {Alpher}\ \emph {et~al.}(1948)\citenamefont {Alpher}, \citenamefont {Bethe},\ and\ \citenamefont {Gamow}}]{Alpher:1948ve}%
  \BibitemOpen
  \bibfield  {author} {\bibinfo {author} {\bibfnamefont {R.~A.}\ \bibnamefont {Alpher}}, \bibinfo {author} {\bibfnamefont {H.}~\bibnamefont {Bethe}},\ and\ \bibinfo {author} {\bibfnamefont {G.}~\bibnamefont {Gamow}},\ }\bibfield  {title} {\bibinfo {title} {{The origin of chemical elements}},\ }\href {https://doi.org/10.1103/PhysRev.73.803} {\bibfield  {journal} {\bibinfo  {journal} {Phys. Rev.}\ }\textbf {\bibinfo {volume} {73}},\ \bibinfo {pages} {803} (\bibinfo {year} {1948})}\BibitemShut {NoStop}%
\bibitem [{\citenamefont {Walker}\ \emph {et~al.}(1991)\citenamefont {Walker}, \citenamefont {Steigman}, \citenamefont {Schramm}, \citenamefont {Olive},\ and\ \citenamefont {Kang}}]{Walker:1991ap}%
  \BibitemOpen
  \bibfield  {author} {\bibinfo {author} {\bibfnamefont {T.~P.}\ \bibnamefont {Walker}}, \bibinfo {author} {\bibfnamefont {G.}~\bibnamefont {Steigman}}, \bibinfo {author} {\bibfnamefont {D.~N.}\ \bibnamefont {Schramm}}, \bibinfo {author} {\bibfnamefont {K.~A.}\ \bibnamefont {Olive}},\ and\ \bibinfo {author} {\bibfnamefont {H.-S.}\ \bibnamefont {Kang}},\ }\bibfield  {title} {\bibinfo {title} {{Primordial nucleosynthesis redux}},\ }\href {https://doi.org/10.1086/170255} {\bibfield  {journal} {\bibinfo  {journal} {Astrophys. J.}\ }\textbf {\bibinfo {volume} {376}},\ \bibinfo {pages} {51} (\bibinfo {year} {1991})}\BibitemShut {NoStop}%
\bibitem [{\citenamefont {Olive}\ \emph {et~al.}(2000)\citenamefont {Olive}, \citenamefont {Steigman},\ and\ \citenamefont {Walker}}]{Olive:1999ij}%
  \BibitemOpen
  \bibfield  {author} {\bibinfo {author} {\bibfnamefont {K.~A.}\ \bibnamefont {Olive}}, \bibinfo {author} {\bibfnamefont {G.}~\bibnamefont {Steigman}},\ and\ \bibinfo {author} {\bibfnamefont {T.~P.}\ \bibnamefont {Walker}},\ }\bibfield  {title} {\bibinfo {title} {{Primordial nucleosynthesis: Theory and observations}},\ }\href {https://doi.org/10.1016/S0370-1573(00)00031-4} {\bibfield  {journal} {\bibinfo  {journal} {Phys. Rept.}\ }\textbf {\bibinfo {volume} {333}},\ \bibinfo {pages} {389} (\bibinfo {year} {2000})},\ \Eprint {https://arxiv.org/abs/astro-ph/9905320} {arXiv:astro-ph/9905320} \BibitemShut {NoStop}%
\bibitem [{\citenamefont {Akrami}\ \emph {et~al.}(2020)\citenamefont {Akrami} \emph {et~al.}}]{Akrami2020}%
  \BibitemOpen
  \bibfield  {author} {\bibinfo {author} {\bibfnamefont {Y.}~\bibnamefont {Akrami}} \emph {et~al.} (\bibinfo {collaboration} {Planck}),\ }\bibfield  {title} {\bibinfo {title} {{Planck 2018 results. X. Constraints on inflation}},\ }\href {https://doi.org/10.1051/0004-6361/201833887} {\bibfield  {journal} {\bibinfo  {journal} {Astron. Astrophys.}\ }\textbf {\bibinfo {volume} {641}},\ \bibinfo {pages} {A10} (\bibinfo {year} {2020})},\ \Eprint {https://arxiv.org/abs/1807.06211} {arXiv:1807.06211 [astro-ph.CO]} \BibitemShut {NoStop}%
\bibitem [{\citenamefont {Hall}\ \emph {et~al.}(2010)\citenamefont {Hall}, \citenamefont {Jedamzik}, \citenamefont {March-Russell},\ and\ \citenamefont {West}}]{Hall:2009bx}%
  \BibitemOpen
  \bibfield  {author} {\bibinfo {author} {\bibfnamefont {L.~J.}\ \bibnamefont {Hall}}, \bibinfo {author} {\bibfnamefont {K.}~\bibnamefont {Jedamzik}}, \bibinfo {author} {\bibfnamefont {J.}~\bibnamefont {March-Russell}},\ and\ \bibinfo {author} {\bibfnamefont {S.~M.}\ \bibnamefont {West}},\ }\bibfield  {title} {\bibinfo {title} {{Freeze-In Production of FIMP Dark Matter}},\ }\href {https://doi.org/10.1007/JHEP03(2010)080} {\bibfield  {journal} {\bibinfo  {journal} {JHEP}\ }\textbf {\bibinfo {volume} {03}},\ \bibinfo {pages} {080}},\ \Eprint {https://arxiv.org/abs/0911.1120} {arXiv:0911.1120 [hep-ph]} \BibitemShut {NoStop}%
\bibitem [{\citenamefont {Chu}\ \emph {et~al.}(2012)\citenamefont {Chu}, \citenamefont {Hambye},\ and\ \citenamefont {Tytgat}}]{Chu:2011be}%
  \BibitemOpen
  \bibfield  {author} {\bibinfo {author} {\bibfnamefont {X.}~\bibnamefont {Chu}}, \bibinfo {author} {\bibfnamefont {T.}~\bibnamefont {Hambye}},\ and\ \bibinfo {author} {\bibfnamefont {M.~H.~G.}\ \bibnamefont {Tytgat}},\ }\bibfield  {title} {\bibinfo {title} {{The Four Basic Ways of Creating Dark Matter Through a Portal}},\ }\href {https://doi.org/10.1088/1475-7516/2012/05/034} {\bibfield  {journal} {\bibinfo  {journal} {JCAP}\ }\textbf {\bibinfo {volume} {05}},\ \bibinfo {pages} {034}},\ \Eprint {https://arxiv.org/abs/1112.0493} {arXiv:1112.0493 [hep-ph]} \BibitemShut {NoStop}%
\bibitem [{\citenamefont {Mambrini}\ \emph {et~al.}(2013)\citenamefont {Mambrini}, \citenamefont {Olive}, \citenamefont {Quevillon},\ and\ \citenamefont {Zaldivar}}]{Mambrini:2013iaa}%
  \BibitemOpen
  \bibfield  {author} {\bibinfo {author} {\bibfnamefont {Y.}~\bibnamefont {Mambrini}}, \bibinfo {author} {\bibfnamefont {K.~A.}\ \bibnamefont {Olive}}, \bibinfo {author} {\bibfnamefont {J.}~\bibnamefont {Quevillon}},\ and\ \bibinfo {author} {\bibfnamefont {B.}~\bibnamefont {Zaldivar}},\ }\bibfield  {title} {\bibinfo {title} {{Gauge Coupling Unification and Nonequilibrium Thermal Dark Matter}},\ }\href {https://doi.org/10.1103/PhysRevLett.110.241306} {\bibfield  {journal} {\bibinfo  {journal} {Phys. Rev. Lett.}\ }\textbf {\bibinfo {volume} {110}},\ \bibinfo {pages} {241306} (\bibinfo {year} {2013})},\ \Eprint {https://arxiv.org/abs/1302.4438} {arXiv:1302.4438 [hep-ph]} \BibitemShut {NoStop}%
\bibitem [{\citenamefont {Chu}\ \emph {et~al.}(2014)\citenamefont {Chu}, \citenamefont {Mambrini}, \citenamefont {Quevillon},\ and\ \citenamefont {Zaldivar}}]{Chu:2013jja}%
  \BibitemOpen
  \bibfield  {author} {\bibinfo {author} {\bibfnamefont {X.}~\bibnamefont {Chu}}, \bibinfo {author} {\bibfnamefont {Y.}~\bibnamefont {Mambrini}}, \bibinfo {author} {\bibfnamefont {J.}~\bibnamefont {Quevillon}},\ and\ \bibinfo {author} {\bibfnamefont {B.}~\bibnamefont {Zaldivar}},\ }\bibfield  {title} {\bibinfo {title} {{Thermal and non-thermal production of dark matter via Z'-portal(s)}},\ }\href {https://doi.org/10.1088/1475-7516/2014/01/034} {\bibfield  {journal} {\bibinfo  {journal} {JCAP}\ }\textbf {\bibinfo {volume} {01}},\ \bibinfo {pages} {034}},\ \Eprint {https://arxiv.org/abs/1306.4677} {arXiv:1306.4677 [hep-ph]} \BibitemShut {NoStop}%
\bibitem [{\citenamefont {Kaneta}\ \emph {et~al.}(2017{\natexlab{a}})\citenamefont {Kaneta}, \citenamefont {Lee},\ and\ \citenamefont {Yun}}]{Kaneta:2016wvf}%
  \BibitemOpen
  \bibfield  {author} {\bibinfo {author} {\bibfnamefont {K.}~\bibnamefont {Kaneta}}, \bibinfo {author} {\bibfnamefont {H.-S.}\ \bibnamefont {Lee}},\ and\ \bibinfo {author} {\bibfnamefont {S.}~\bibnamefont {Yun}},\ }\bibfield  {title} {\bibinfo {title} {{Portal Connecting Dark Photons and Axions}},\ }\href {https://doi.org/10.1103/PhysRevLett.118.101802} {\bibfield  {journal} {\bibinfo  {journal} {Phys. Rev. Lett.}\ }\textbf {\bibinfo {volume} {118}},\ \bibinfo {pages} {101802} (\bibinfo {year} {2017}{\natexlab{a}})},\ \Eprint {https://arxiv.org/abs/1611.01466} {arXiv:1611.01466 [hep-ph]} \BibitemShut {NoStop}%
\bibitem [{\citenamefont {Kaneta}\ \emph {et~al.}(2017{\natexlab{b}})\citenamefont {Kaneta}, \citenamefont {Lee},\ and\ \citenamefont {Yun}}]{Kaneta:2017wfh}%
  \BibitemOpen
  \bibfield  {author} {\bibinfo {author} {\bibfnamefont {K.}~\bibnamefont {Kaneta}}, \bibinfo {author} {\bibfnamefont {H.-S.}\ \bibnamefont {Lee}},\ and\ \bibinfo {author} {\bibfnamefont {S.}~\bibnamefont {Yun}},\ }\bibfield  {title} {\bibinfo {title} {{Dark photon relic dark matter production through the dark axion portal}},\ }\href {https://doi.org/10.1103/PhysRevD.95.115032} {\bibfield  {journal} {\bibinfo  {journal} {Phys. Rev. D}\ }\textbf {\bibinfo {volume} {95}},\ \bibinfo {pages} {115032} (\bibinfo {year} {2017}{\natexlab{b}})},\ \Eprint {https://arxiv.org/abs/1704.07542} {arXiv:1704.07542 [hep-ph]} \BibitemShut {NoStop}%
\bibitem [{\citenamefont {Bernal}\ \emph {et~al.}(2017)\citenamefont {Bernal}, \citenamefont {Heikinheimo}, \citenamefont {Tenkanen}, \citenamefont {Tuominen},\ and\ \citenamefont {Vaskonen}}]{Bernal:2017kxu}%
  \BibitemOpen
  \bibfield  {author} {\bibinfo {author} {\bibfnamefont {N.}~\bibnamefont {Bernal}}, \bibinfo {author} {\bibfnamefont {M.}~\bibnamefont {Heikinheimo}}, \bibinfo {author} {\bibfnamefont {T.}~\bibnamefont {Tenkanen}}, \bibinfo {author} {\bibfnamefont {K.}~\bibnamefont {Tuominen}},\ and\ \bibinfo {author} {\bibfnamefont {V.}~\bibnamefont {Vaskonen}},\ }\bibfield  {title} {\bibinfo {title} {{The Dawn of FIMP Dark Matter: A Review of Models and Constraints}},\ }\href {https://doi.org/10.1142/S0217751X1730023X} {\bibfield  {journal} {\bibinfo  {journal} {Int. J. Mod. Phys. A}\ }\textbf {\bibinfo {volume} {32}},\ \bibinfo {pages} {1730023} (\bibinfo {year} {2017})},\ \Eprint {https://arxiv.org/abs/1706.07442} {arXiv:1706.07442 [hep-ph]} \BibitemShut {NoStop}%
\bibitem [{\citenamefont {Biswas}\ \emph {et~al.}(2019)\citenamefont {Biswas}, \citenamefont {Borah},\ and\ \citenamefont {Dasgupta}}]{Biswas:2018aib}%
  \BibitemOpen
  \bibfield  {author} {\bibinfo {author} {\bibfnamefont {A.}~\bibnamefont {Biswas}}, \bibinfo {author} {\bibfnamefont {D.}~\bibnamefont {Borah}},\ and\ \bibinfo {author} {\bibfnamefont {A.}~\bibnamefont {Dasgupta}},\ }\bibfield  {title} {\bibinfo {title} {{UV complete framework of freeze-in massive particle dark matter}},\ }\href {https://doi.org/10.1103/PhysRevD.99.015033} {\bibfield  {journal} {\bibinfo  {journal} {Phys. Rev. D}\ }\textbf {\bibinfo {volume} {99}},\ \bibinfo {pages} {015033} (\bibinfo {year} {2019})},\ \Eprint {https://arxiv.org/abs/1805.06903} {arXiv:1805.06903 [hep-ph]} \BibitemShut {NoStop}%
\bibitem [{\citenamefont {Kaneta}\ \emph {et~al.}(2019)\citenamefont {Kaneta}, \citenamefont {Mambrini},\ and\ \citenamefont {Olive}}]{Kaneta:2019zgw}%
  \BibitemOpen
  \bibfield  {author} {\bibinfo {author} {\bibfnamefont {K.}~\bibnamefont {Kaneta}}, \bibinfo {author} {\bibfnamefont {Y.}~\bibnamefont {Mambrini}},\ and\ \bibinfo {author} {\bibfnamefont {K.~A.}\ \bibnamefont {Olive}},\ }\bibfield  {title} {\bibinfo {title} {{Radiative production of nonthermal dark matter}},\ }\href {https://doi.org/10.1103/PhysRevD.99.063508} {\bibfield  {journal} {\bibinfo  {journal} {Phys. Rev. D}\ }\textbf {\bibinfo {volume} {99}},\ \bibinfo {pages} {063508} (\bibinfo {year} {2019})},\ \Eprint {https://arxiv.org/abs/1901.04449} {arXiv:1901.04449 [hep-ph]} \BibitemShut {NoStop}%
\bibitem [{\citenamefont {Bernal}\ \emph {et~al.}(2019)\citenamefont {Bernal}, \citenamefont {Elahi}, \citenamefont {Maldonado},\ and\ \citenamefont {Unwin}}]{Bernal:2019mhf}%
  \BibitemOpen
  \bibfield  {author} {\bibinfo {author} {\bibfnamefont {N.}~\bibnamefont {Bernal}}, \bibinfo {author} {\bibfnamefont {F.}~\bibnamefont {Elahi}}, \bibinfo {author} {\bibfnamefont {C.}~\bibnamefont {Maldonado}},\ and\ \bibinfo {author} {\bibfnamefont {J.}~\bibnamefont {Unwin}},\ }\bibfield  {title} {\bibinfo {title} {{Ultraviolet Freeze-in and Non-Standard Cosmologies}},\ }\href {https://doi.org/10.1088/1475-7516/2019/11/026} {\bibfield  {journal} {\bibinfo  {journal} {JCAP}\ }\textbf {\bibinfo {volume} {11}},\ \bibinfo {pages} {026}},\ \Eprint {https://arxiv.org/abs/1909.07992} {arXiv:1909.07992 [hep-ph]} \BibitemShut {NoStop}%
\bibitem [{\citenamefont {Bernal}(2020)}]{Bernal:2020gzm}%
  \BibitemOpen
  \bibfield  {author} {\bibinfo {author} {\bibfnamefont {N.}~\bibnamefont {Bernal}},\ }\bibfield  {title} {\bibinfo {title} {{Boosting Freeze-in through Thermalization}},\ }\href {https://doi.org/10.1088/1475-7516/2020/10/006} {\bibfield  {journal} {\bibinfo  {journal} {JCAP}\ }\textbf {\bibinfo {volume} {10}},\ \bibinfo {pages} {006}},\ \Eprint {https://arxiv.org/abs/2005.08988} {arXiv:2005.08988 [hep-ph]} \BibitemShut {NoStop}%
\bibitem [{\citenamefont {Bernal}\ \emph {et~al.}(2020)\citenamefont {Bernal}, \citenamefont {Rubio},\ and\ \citenamefont {Veerm\"ae}}]{Bernal:2020qyu}%
  \BibitemOpen
  \bibfield  {author} {\bibinfo {author} {\bibfnamefont {N.}~\bibnamefont {Bernal}}, \bibinfo {author} {\bibfnamefont {J.}~\bibnamefont {Rubio}},\ and\ \bibinfo {author} {\bibfnamefont {H.}~\bibnamefont {Veerm\"ae}},\ }\bibfield  {title} {\bibinfo {title} {{UV Freeze-in in Starobinsky Inflation}},\ }\href {https://doi.org/10.1088/1475-7516/2020/10/021} {\bibfield  {journal} {\bibinfo  {journal} {JCAP}\ }\textbf {\bibinfo {volume} {10}},\ \bibinfo {pages} {021}},\ \Eprint {https://arxiv.org/abs/2006.02442} {arXiv:2006.02442 [hep-ph]} \BibitemShut {NoStop}%
\bibitem [{\citenamefont {Anastasopoulos}\ \emph {et~al.}(2020)\citenamefont {Anastasopoulos}, \citenamefont {Kaneta}, \citenamefont {Mambrini},\ and\ \citenamefont {Pierre}}]{Anastasopoulos:2020gbu}%
  \BibitemOpen
  \bibfield  {author} {\bibinfo {author} {\bibfnamefont {P.}~\bibnamefont {Anastasopoulos}}, \bibinfo {author} {\bibfnamefont {K.}~\bibnamefont {Kaneta}}, \bibinfo {author} {\bibfnamefont {Y.}~\bibnamefont {Mambrini}},\ and\ \bibinfo {author} {\bibfnamefont {M.}~\bibnamefont {Pierre}},\ }\bibfield  {title} {\bibinfo {title} {{Energy-momentum portal to dark matter and emergent gravity}},\ }\href {https://doi.org/10.1103/PhysRevD.102.055019} {\bibfield  {journal} {\bibinfo  {journal} {Phys. Rev. D}\ }\textbf {\bibinfo {volume} {102}},\ \bibinfo {pages} {055019} (\bibinfo {year} {2020})},\ \Eprint {https://arxiv.org/abs/2007.06534} {arXiv:2007.06534 [hep-ph]} \BibitemShut {NoStop}%
\bibitem [{\citenamefont {Brax}\ \emph {et~al.}(2021{\natexlab{a}})\citenamefont {Brax}, \citenamefont {Kaneta}, \citenamefont {Mambrini},\ and\ \citenamefont {Pierre}}]{Brax:2020gqg}%
  \BibitemOpen
  \bibfield  {author} {\bibinfo {author} {\bibfnamefont {P.}~\bibnamefont {Brax}}, \bibinfo {author} {\bibfnamefont {K.}~\bibnamefont {Kaneta}}, \bibinfo {author} {\bibfnamefont {Y.}~\bibnamefont {Mambrini}},\ and\ \bibinfo {author} {\bibfnamefont {M.}~\bibnamefont {Pierre}},\ }\bibfield  {title} {\bibinfo {title} {{Disformal dark matter}},\ }\href {https://doi.org/10.1103/PhysRevD.103.015028} {\bibfield  {journal} {\bibinfo  {journal} {Phys. Rev. D}\ }\textbf {\bibinfo {volume} {103}},\ \bibinfo {pages} {015028} (\bibinfo {year} {2021}{\natexlab{a}})},\ \Eprint {https://arxiv.org/abs/2011.11647} {arXiv:2011.11647 [hep-ph]} \BibitemShut {NoStop}%
\bibitem [{\citenamefont {Brax}\ \emph {et~al.}(2021{\natexlab{b}})\citenamefont {Brax}, \citenamefont {Kaneta}, \citenamefont {Mambrini},\ and\ \citenamefont {Pierre}}]{Brax:2021gpe}%
  \BibitemOpen
  \bibfield  {author} {\bibinfo {author} {\bibfnamefont {P.}~\bibnamefont {Brax}}, \bibinfo {author} {\bibfnamefont {K.}~\bibnamefont {Kaneta}}, \bibinfo {author} {\bibfnamefont {Y.}~\bibnamefont {Mambrini}},\ and\ \bibinfo {author} {\bibfnamefont {M.}~\bibnamefont {Pierre}},\ }\bibfield  {title} {\bibinfo {title} {{Metastable Conformal Dark Matter}},\ }\href {https://doi.org/10.1103/PhysRevD.103.115016} {\bibfield  {journal} {\bibinfo  {journal} {Phys. Rev. D}\ }\textbf {\bibinfo {volume} {103}},\ \bibinfo {pages} {115016} (\bibinfo {year} {2021}{\natexlab{b}})},\ \Eprint {https://arxiv.org/abs/2103.02615} {arXiv:2103.02615 [hep-ph]} \BibitemShut {NoStop}%
\bibitem [{\citenamefont {Kaneta}\ \emph {et~al.}(2021)\citenamefont {Kaneta}, \citenamefont {Ko},\ and\ \citenamefont {Park}}]{Kaneta:2021pyx}%
  \BibitemOpen
  \bibfield  {author} {\bibinfo {author} {\bibfnamefont {K.}~\bibnamefont {Kaneta}}, \bibinfo {author} {\bibfnamefont {P.}~\bibnamefont {Ko}},\ and\ \bibinfo {author} {\bibfnamefont {W.-I.}\ \bibnamefont {Park}},\ }\bibfield  {title} {\bibinfo {title} {{Conformal portal to dark matter}},\ }\href {https://doi.org/10.1103/PhysRevD.104.075018} {\bibfield  {journal} {\bibinfo  {journal} {Phys. Rev. D}\ }\textbf {\bibinfo {volume} {104}},\ \bibinfo {pages} {075018} (\bibinfo {year} {2021})},\ \Eprint {https://arxiv.org/abs/2106.01923} {arXiv:2106.01923 [hep-ph]} \BibitemShut {NoStop}%
\bibitem [{\citenamefont {Ghosh}\ and\ \citenamefont {Mukhopadhyay}(2022)}]{Ghosh:2022hen}%
  \BibitemOpen
  \bibfield  {author} {\bibinfo {author} {\bibfnamefont {A.}~\bibnamefont {Ghosh}}\ and\ \bibinfo {author} {\bibfnamefont {S.}~\bibnamefont {Mukhopadhyay}},\ }\bibfield  {title} {\bibinfo {title} {{Momentum distribution of dark matter produced in inflaton decay: Effect of inflaton mediated scatterings}},\ }\href {https://doi.org/10.1103/PhysRevD.106.043519} {\bibfield  {journal} {\bibinfo  {journal} {Phys. Rev. D}\ }\textbf {\bibinfo {volume} {106}},\ \bibinfo {pages} {043519} (\bibinfo {year} {2022})},\ \Eprint {https://arxiv.org/abs/2205.03440} {arXiv:2205.03440 [hep-ph]} \BibitemShut {NoStop}%
\bibitem [{\citenamefont {Giudice}\ \emph {et~al.}(1999)\citenamefont {Giudice}, \citenamefont {Peloso}, \citenamefont {Riotto},\ and\ \citenamefont {Tkachev}}]{Giudice:1999fb}%
  \BibitemOpen
  \bibfield  {author} {\bibinfo {author} {\bibfnamefont {G.~F.}\ \bibnamefont {Giudice}}, \bibinfo {author} {\bibfnamefont {M.}~\bibnamefont {Peloso}}, \bibinfo {author} {\bibfnamefont {A.}~\bibnamefont {Riotto}},\ and\ \bibinfo {author} {\bibfnamefont {I.}~\bibnamefont {Tkachev}},\ }\bibfield  {title} {\bibinfo {title} {{Production of massive fermions at preheating and leptogenesis}},\ }\href {https://doi.org/10.1088/1126-6708/1999/08/014} {\bibfield  {journal} {\bibinfo  {journal} {JHEP}\ }\textbf {\bibinfo {volume} {08}},\ \bibinfo {pages} {014}},\ \Eprint {https://arxiv.org/abs/hep-ph/9905242} {arXiv:hep-ph/9905242} \BibitemShut {NoStop}%
\bibitem [{\citenamefont {Asaka}\ \emph {et~al.}(1999)\citenamefont {Asaka}, \citenamefont {Hamaguchi}, \citenamefont {Kawasaki},\ and\ \citenamefont {Yanagida}}]{Asaka:1999yd}%
  \BibitemOpen
  \bibfield  {author} {\bibinfo {author} {\bibfnamefont {T.}~\bibnamefont {Asaka}}, \bibinfo {author} {\bibfnamefont {K.}~\bibnamefont {Hamaguchi}}, \bibinfo {author} {\bibfnamefont {M.}~\bibnamefont {Kawasaki}},\ and\ \bibinfo {author} {\bibfnamefont {T.}~\bibnamefont {Yanagida}},\ }\bibfield  {title} {\bibinfo {title} {{Leptogenesis in inflaton decay}},\ }\href {https://doi.org/10.1016/S0370-2693(99)01020-5} {\bibfield  {journal} {\bibinfo  {journal} {Phys. Lett. B}\ }\textbf {\bibinfo {volume} {464}},\ \bibinfo {pages} {12} (\bibinfo {year} {1999})},\ \Eprint {https://arxiv.org/abs/hep-ph/9906366} {arXiv:hep-ph/9906366} \BibitemShut {NoStop}%
\bibitem [{\citenamefont {Kaneta}\ \emph {et~al.}(2020)\citenamefont {Kaneta}, \citenamefont {Mambrini}, \citenamefont {Olive},\ and\ \citenamefont {Verner}}]{Kaneta:2019yjn}%
  \BibitemOpen
  \bibfield  {author} {\bibinfo {author} {\bibfnamefont {K.}~\bibnamefont {Kaneta}}, \bibinfo {author} {\bibfnamefont {Y.}~\bibnamefont {Mambrini}}, \bibinfo {author} {\bibfnamefont {K.~A.}\ \bibnamefont {Olive}},\ and\ \bibinfo {author} {\bibfnamefont {S.}~\bibnamefont {Verner}},\ }\bibfield  {title} {\bibinfo {title} {{Inflation and Leptogenesis in High-Scale Supersymmetry}},\ }\href {https://doi.org/10.1103/PhysRevD.101.015002} {\bibfield  {journal} {\bibinfo  {journal} {Phys. Rev. D}\ }\textbf {\bibinfo {volume} {101}},\ \bibinfo {pages} {015002} (\bibinfo {year} {2020})},\ \Eprint {https://arxiv.org/abs/1911.02463} {arXiv:1911.02463 [hep-ph]} \BibitemShut {NoStop}%
\bibitem [{\citenamefont {Ema}\ \emph {et~al.}(2020)\citenamefont {Ema}, \citenamefont {Jinno},\ and\ \citenamefont {Nakayama}}]{Ema:2020ggo}%
  \BibitemOpen
  \bibfield  {author} {\bibinfo {author} {\bibfnamefont {Y.}~\bibnamefont {Ema}}, \bibinfo {author} {\bibfnamefont {R.}~\bibnamefont {Jinno}},\ and\ \bibinfo {author} {\bibfnamefont {K.}~\bibnamefont {Nakayama}},\ }\bibfield  {title} {\bibinfo {title} {{High-frequency Graviton from Inflaton Oscillation}},\ }\href {https://doi.org/10.1088/1475-7516/2020/09/015} {\bibfield  {journal} {\bibinfo  {journal} {JCAP}\ }\textbf {\bibinfo {volume} {09}},\ \bibinfo {pages} {015}},\ \Eprint {https://arxiv.org/abs/2006.09972} {arXiv:2006.09972 [astro-ph.CO]} \BibitemShut {NoStop}%
\bibitem [{\citenamefont {Klose}\ \emph {et~al.}(2022)\citenamefont {Klose}, \citenamefont {Laine},\ and\ \citenamefont {Procacci}}]{Klose:2022knn}%
  \BibitemOpen
  \bibfield  {author} {\bibinfo {author} {\bibfnamefont {P.}~\bibnamefont {Klose}}, \bibinfo {author} {\bibfnamefont {M.}~\bibnamefont {Laine}},\ and\ \bibinfo {author} {\bibfnamefont {S.}~\bibnamefont {Procacci}},\ }\bibfield  {title} {\bibinfo {title} {{Gravitational wave background from non-Abelian reheating after axion-like inflation}},\ }\href {https://doi.org/10.1088/1475-7516/2022/05/021} {\bibfield  {journal} {\bibinfo  {journal} {JCAP}\ }\textbf {\bibinfo {volume} {05}},\ \bibinfo {pages} {021}},\ \Eprint {https://arxiv.org/abs/2201.02317} {arXiv:2201.02317 [hep-ph]} \BibitemShut {NoStop}%
\bibitem [{\citenamefont {Nakayama}\ and\ \citenamefont {Tang}(2019)}]{Nakayama2019}%
  \BibitemOpen
  \bibfield  {author} {\bibinfo {author} {\bibfnamefont {K.}~\bibnamefont {Nakayama}}\ and\ \bibinfo {author} {\bibfnamefont {Y.}~\bibnamefont {Tang}},\ }\bibfield  {title} {\bibinfo {title} {{Stochastic Gravitational Waves from Particle Origin}},\ }\href {https://doi.org/10.1016/j.physletb.2018.11.023} {\bibfield  {journal} {\bibinfo  {journal} {Phys. Lett. B}\ }\textbf {\bibinfo {volume} {788}},\ \bibinfo {pages} {341} (\bibinfo {year} {2019})},\ \Eprint {https://arxiv.org/abs/1810.04975} {arXiv:1810.04975 [hep-ph]} \BibitemShut {NoStop}%
\bibitem [{\citenamefont {Barman}\ \emph {et~al.}(2023{\natexlab{a}})\citenamefont {Barman}, \citenamefont {Bernal}, \citenamefont {Xu},\ and\ \citenamefont {Zapata}}]{Barman2023}%
  \BibitemOpen
  \bibfield  {author} {\bibinfo {author} {\bibfnamefont {B.}~\bibnamefont {Barman}}, \bibinfo {author} {\bibfnamefont {N.}~\bibnamefont {Bernal}}, \bibinfo {author} {\bibfnamefont {Y.}~\bibnamefont {Xu}},\ and\ \bibinfo {author} {\bibfnamefont {O.}~\bibnamefont {Zapata}},\ }\bibfield  {title} {\bibinfo {title} {{Gravitational wave from graviton Bremsstrahlung during reheating}},\ }\href {https://doi.org/10.1088/1475-7516/2023/05/019} {\bibfield  {journal} {\bibinfo  {journal} {JCAP}\ }\textbf {\bibinfo {volume} {05}},\ \bibinfo {pages} {019}},\ \Eprint {https://arxiv.org/abs/2301.11345} {arXiv:2301.11345 [hep-ph]} \BibitemShut {NoStop}%
\bibitem [{\citenamefont {Barman}\ \emph {et~al.}(2023{\natexlab{b}})\citenamefont {Barman}, \citenamefont {Bernal}, \citenamefont {Xu},\ and\ \citenamefont {Zapata}}]{Barman:2023rpg}%
  \BibitemOpen
  \bibfield  {author} {\bibinfo {author} {\bibfnamefont {B.}~\bibnamefont {Barman}}, \bibinfo {author} {\bibfnamefont {N.}~\bibnamefont {Bernal}}, \bibinfo {author} {\bibfnamefont {Y.}~\bibnamefont {Xu}},\ and\ \bibinfo {author} {\bibfnamefont {O.}~\bibnamefont {Zapata}},\ }\bibfield  {title} {\bibinfo {title} {{Bremsstrahlung-induced Gravitational Waves in Monomial Potentials during Reheating}},\ }\href@noop {} {\  (\bibinfo {year} {2023}{\natexlab{b}})},\ \Eprint {https://arxiv.org/abs/2305.16388} {arXiv:2305.16388 [hep-ph]} \BibitemShut {NoStop}%
\bibitem [{\citenamefont {Apreda}\ \emph {et~al.}(2002)\citenamefont {Apreda}, \citenamefont {Maggiore}, \citenamefont {Nicolis},\ and\ \citenamefont {Riotto}}]{Apreda:2001us}%
  \BibitemOpen
  \bibfield  {author} {\bibinfo {author} {\bibfnamefont {R.}~\bibnamefont {Apreda}}, \bibinfo {author} {\bibfnamefont {M.}~\bibnamefont {Maggiore}}, \bibinfo {author} {\bibfnamefont {A.}~\bibnamefont {Nicolis}},\ and\ \bibinfo {author} {\bibfnamefont {A.}~\bibnamefont {Riotto}},\ }\bibfield  {title} {\bibinfo {title} {{Gravitational waves from electroweak phase transitions}},\ }\href {https://doi.org/10.1016/S0550-3213(02)00264-X} {\bibfield  {journal} {\bibinfo  {journal} {Nucl. Phys. B}\ }\textbf {\bibinfo {volume} {631}},\ \bibinfo {pages} {342} (\bibinfo {year} {2002})},\ \Eprint {https://arxiv.org/abs/gr-qc/0107033} {arXiv:gr-qc/0107033} \BibitemShut {NoStop}%
\bibitem [{\citenamefont {Grojean}\ and\ \citenamefont {Servant}(2007)}]{Grojean:2006bp}%
  \BibitemOpen
  \bibfield  {author} {\bibinfo {author} {\bibfnamefont {C.}~\bibnamefont {Grojean}}\ and\ \bibinfo {author} {\bibfnamefont {G.}~\bibnamefont {Servant}},\ }\bibfield  {title} {\bibinfo {title} {{Gravitational Waves from Phase Transitions at the Electroweak Scale and Beyond}},\ }\href {https://doi.org/10.1103/PhysRevD.75.043507} {\bibfield  {journal} {\bibinfo  {journal} {Phys. Rev. D}\ }\textbf {\bibinfo {volume} {75}},\ \bibinfo {pages} {043507} (\bibinfo {year} {2007})},\ \Eprint {https://arxiv.org/abs/hep-ph/0607107} {arXiv:hep-ph/0607107} \BibitemShut {NoStop}%
\bibitem [{\citenamefont {Espinosa}\ \emph {et~al.}(2008)\citenamefont {Espinosa}, \citenamefont {Konstandin}, \citenamefont {No},\ and\ \citenamefont {Quiros}}]{Espinosa:2008kw}%
  \BibitemOpen
  \bibfield  {author} {\bibinfo {author} {\bibfnamefont {J.~R.}\ \bibnamefont {Espinosa}}, \bibinfo {author} {\bibfnamefont {T.}~\bibnamefont {Konstandin}}, \bibinfo {author} {\bibfnamefont {J.~M.}\ \bibnamefont {No}},\ and\ \bibinfo {author} {\bibfnamefont {M.}~\bibnamefont {Quiros}},\ }\bibfield  {title} {\bibinfo {title} {{Some Cosmological Implications of Hidden Sectors}},\ }\href {https://doi.org/10.1103/PhysRevD.78.123528} {\bibfield  {journal} {\bibinfo  {journal} {Phys. Rev. D}\ }\textbf {\bibinfo {volume} {78}},\ \bibinfo {pages} {123528} (\bibinfo {year} {2008})},\ \Eprint {https://arxiv.org/abs/0809.3215} {arXiv:0809.3215 [hep-ph]} \BibitemShut {NoStop}%
\bibitem [{\citenamefont {Caprini}\ \emph {et~al.}(2016)\citenamefont {Caprini} \emph {et~al.}}]{Caprini:2015zlo}%
  \BibitemOpen
  \bibfield  {author} {\bibinfo {author} {\bibfnamefont {C.}~\bibnamefont {Caprini}} \emph {et~al.},\ }\bibfield  {title} {\bibinfo {title} {{Science with the space-based interferometer eLISA. II: Gravitational waves from cosmological phase transitions}},\ }\href {https://doi.org/10.1088/1475-7516/2016/04/001} {\bibfield  {journal} {\bibinfo  {journal} {JCAP}\ }\textbf {\bibinfo {volume} {04}},\ \bibinfo {pages} {001}},\ \Eprint {https://arxiv.org/abs/1512.06239} {arXiv:1512.06239 [astro-ph.CO]} \BibitemShut {NoStop}%
\bibitem [{\citenamefont {Kakizaki}\ \emph {et~al.}(2015)\citenamefont {Kakizaki}, \citenamefont {Kanemura},\ and\ \citenamefont {Matsui}}]{Kakizaki:2015wua}%
  \BibitemOpen
  \bibfield  {author} {\bibinfo {author} {\bibfnamefont {M.}~\bibnamefont {Kakizaki}}, \bibinfo {author} {\bibfnamefont {S.}~\bibnamefont {Kanemura}},\ and\ \bibinfo {author} {\bibfnamefont {T.}~\bibnamefont {Matsui}},\ }\bibfield  {title} {\bibinfo {title} {{Gravitational waves as a probe of extended scalar sectors with the first order electroweak phase transition}},\ }\href {https://doi.org/10.1103/PhysRevD.92.115007} {\bibfield  {journal} {\bibinfo  {journal} {Phys. Rev. D}\ }\textbf {\bibinfo {volume} {92}},\ \bibinfo {pages} {115007} (\bibinfo {year} {2015})},\ \Eprint {https://arxiv.org/abs/1509.08394} {arXiv:1509.08394 [hep-ph]} \BibitemShut {NoStop}%
\bibitem [{\citenamefont {Hashino}\ \emph {et~al.}(2016)\citenamefont {Hashino}, \citenamefont {Kakizaki}, \citenamefont {Kanemura},\ and\ \citenamefont {Matsui}}]{Hashino:2016rvx}%
  \BibitemOpen
  \bibfield  {author} {\bibinfo {author} {\bibfnamefont {K.}~\bibnamefont {Hashino}}, \bibinfo {author} {\bibfnamefont {M.}~\bibnamefont {Kakizaki}}, \bibinfo {author} {\bibfnamefont {S.}~\bibnamefont {Kanemura}},\ and\ \bibinfo {author} {\bibfnamefont {T.}~\bibnamefont {Matsui}},\ }\bibfield  {title} {\bibinfo {title} {{Synergy between measurements of gravitational waves and the triple-Higgs coupling in probing the first-order electroweak phase transition}},\ }\href {https://doi.org/10.1103/PhysRevD.94.015005} {\bibfield  {journal} {\bibinfo  {journal} {Phys. Rev. D}\ }\textbf {\bibinfo {volume} {94}},\ \bibinfo {pages} {015005} (\bibinfo {year} {2016})},\ \Eprint {https://arxiv.org/abs/1604.02069} {arXiv:1604.02069 [hep-ph]} \BibitemShut {NoStop}%
\bibitem [{\citenamefont {Hashino}\ \emph {et~al.}(2017)\citenamefont {Hashino}, \citenamefont {Kakizaki}, \citenamefont {Kanemura}, \citenamefont {Ko},\ and\ \citenamefont {Matsui}}]{Hashino:2016xoj}%
  \BibitemOpen
  \bibfield  {author} {\bibinfo {author} {\bibfnamefont {K.}~\bibnamefont {Hashino}}, \bibinfo {author} {\bibfnamefont {M.}~\bibnamefont {Kakizaki}}, \bibinfo {author} {\bibfnamefont {S.}~\bibnamefont {Kanemura}}, \bibinfo {author} {\bibfnamefont {P.}~\bibnamefont {Ko}},\ and\ \bibinfo {author} {\bibfnamefont {T.}~\bibnamefont {Matsui}},\ }\bibfield  {title} {\bibinfo {title} {{Gravitational waves and Higgs boson couplings for exploring first order phase transition in the model with a singlet scalar field}},\ }\href {https://doi.org/10.1016/j.physletb.2016.12.052} {\bibfield  {journal} {\bibinfo  {journal} {Phys. Lett. B}\ }\textbf {\bibinfo {volume} {766}},\ \bibinfo {pages} {49} (\bibinfo {year} {2017})},\ \Eprint {https://arxiv.org/abs/1609.00297} {arXiv:1609.00297 [hep-ph]} \BibitemShut {NoStop}%
\bibitem [{\citenamefont {Athron}\ \emph {et~al.}(2023)\citenamefont {Athron}, \citenamefont {Bal\'azs}, \citenamefont {Fowlie}, \citenamefont {Morris},\ and\ \citenamefont {Wu}}]{Athron:2023xlk}%
  \BibitemOpen
  \bibfield  {author} {\bibinfo {author} {\bibfnamefont {P.}~\bibnamefont {Athron}}, \bibinfo {author} {\bibfnamefont {C.}~\bibnamefont {Bal\'azs}}, \bibinfo {author} {\bibfnamefont {A.}~\bibnamefont {Fowlie}}, \bibinfo {author} {\bibfnamefont {L.}~\bibnamefont {Morris}},\ and\ \bibinfo {author} {\bibfnamefont {L.}~\bibnamefont {Wu}},\ }\bibfield  {title} {\bibinfo {title} {{Cosmological phase transitions: from perturbative particle physics to gravitational waves}},\ }\href@noop {} {\  (\bibinfo {year} {2023})},\ \Eprint {https://arxiv.org/abs/2305.02357} {arXiv:2305.02357 [hep-ph]} \BibitemShut {NoStop}%
\bibitem [{\citenamefont {Vilenkin}(1981)}]{Vilenkin:1981bx}%
  \BibitemOpen
  \bibfield  {author} {\bibinfo {author} {\bibfnamefont {A.}~\bibnamefont {Vilenkin}},\ }\bibfield  {title} {\bibinfo {title} {{Gravitational radiation from cosmic strings}},\ }\href {https://doi.org/10.1016/0370-2693(81)91144-8} {\bibfield  {journal} {\bibinfo  {journal} {Phys. Lett. B}\ }\textbf {\bibinfo {volume} {107}},\ \bibinfo {pages} {47} (\bibinfo {year} {1981})}\BibitemShut {NoStop}%
\bibitem [{\citenamefont {Vachaspati}\ and\ \citenamefont {Vilenkin}(1985)}]{Vachaspati:1984gt}%
  \BibitemOpen
  \bibfield  {author} {\bibinfo {author} {\bibfnamefont {T.}~\bibnamefont {Vachaspati}}\ and\ \bibinfo {author} {\bibfnamefont {A.}~\bibnamefont {Vilenkin}},\ }\bibfield  {title} {\bibinfo {title} {{Gravitational Radiation from Cosmic Strings}},\ }\href {https://doi.org/10.1103/PhysRevD.31.3052} {\bibfield  {journal} {\bibinfo  {journal} {Phys. Rev. D}\ }\textbf {\bibinfo {volume} {31}},\ \bibinfo {pages} {3052} (\bibinfo {year} {1985})}\BibitemShut {NoStop}%
\bibitem [{\citenamefont {Caldwell}\ and\ \citenamefont {Allen}(1992)}]{Caldwell:1991jj}%
  \BibitemOpen
  \bibfield  {author} {\bibinfo {author} {\bibfnamefont {R.~R.}\ \bibnamefont {Caldwell}}\ and\ \bibinfo {author} {\bibfnamefont {B.}~\bibnamefont {Allen}},\ }\bibfield  {title} {\bibinfo {title} {{Cosmological constraints on cosmic string gravitational radiation}},\ }\href {https://doi.org/10.1103/PhysRevD.45.3447} {\bibfield  {journal} {\bibinfo  {journal} {Phys. Rev. D}\ }\textbf {\bibinfo {volume} {45}},\ \bibinfo {pages} {3447} (\bibinfo {year} {1992})}\BibitemShut {NoStop}%
\bibitem [{\citenamefont {Damour}\ and\ \citenamefont {Vilenkin}(2000)}]{Damour:2000wa}%
  \BibitemOpen
  \bibfield  {author} {\bibinfo {author} {\bibfnamefont {T.}~\bibnamefont {Damour}}\ and\ \bibinfo {author} {\bibfnamefont {A.}~\bibnamefont {Vilenkin}},\ }\bibfield  {title} {\bibinfo {title} {{Gravitational wave bursts from cosmic strings}},\ }\href {https://doi.org/10.1103/PhysRevLett.85.3761} {\bibfield  {journal} {\bibinfo  {journal} {Phys. Rev. Lett.}\ }\textbf {\bibinfo {volume} {85}},\ \bibinfo {pages} {3761} (\bibinfo {year} {2000})},\ \Eprint {https://arxiv.org/abs/gr-qc/0004075} {arXiv:gr-qc/0004075} \BibitemShut {NoStop}%
\bibitem [{\citenamefont {Damour}\ and\ \citenamefont {Vilenkin}(2001)}]{Damour:2001bk}%
  \BibitemOpen
  \bibfield  {author} {\bibinfo {author} {\bibfnamefont {T.}~\bibnamefont {Damour}}\ and\ \bibinfo {author} {\bibfnamefont {A.}~\bibnamefont {Vilenkin}},\ }\bibfield  {title} {\bibinfo {title} {{Gravitational wave bursts from cusps and kinks on cosmic strings}},\ }\href {https://doi.org/10.1103/PhysRevD.64.064008} {\bibfield  {journal} {\bibinfo  {journal} {Phys. Rev. D}\ }\textbf {\bibinfo {volume} {64}},\ \bibinfo {pages} {064008} (\bibinfo {year} {2001})},\ \Eprint {https://arxiv.org/abs/gr-qc/0104026} {arXiv:gr-qc/0104026} \BibitemShut {NoStop}%
\bibitem [{\citenamefont {Hiramatsu}\ \emph {et~al.}(2014)\citenamefont {Hiramatsu}, \citenamefont {Kawasaki},\ and\ \citenamefont {Saikawa}}]{Hiramatsu:2013qaa}%
  \BibitemOpen
  \bibfield  {author} {\bibinfo {author} {\bibfnamefont {T.}~\bibnamefont {Hiramatsu}}, \bibinfo {author} {\bibfnamefont {M.}~\bibnamefont {Kawasaki}},\ and\ \bibinfo {author} {\bibfnamefont {K.}~\bibnamefont {Saikawa}},\ }\bibfield  {title} {\bibinfo {title} {{On the estimation of gravitational wave spectrum from cosmic domain walls}},\ }\href {https://doi.org/10.1088/1475-7516/2014/02/031} {\bibfield  {journal} {\bibinfo  {journal} {JCAP}\ }\textbf {\bibinfo {volume} {02}},\ \bibinfo {pages} {031}},\ \Eprint {https://arxiv.org/abs/1309.5001} {arXiv:1309.5001 [astro-ph.CO]} \BibitemShut {NoStop}%
\bibitem [{\citenamefont {Amaro-Seoane}\ \emph {et~al.}(2017)\citenamefont {Amaro-Seoane} \emph {et~al.}}]{LISA:2017pwj}%
  \BibitemOpen
  \bibfield  {author} {\bibinfo {author} {\bibfnamefont {P.}~\bibnamefont {Amaro-Seoane}} \emph {et~al.} (\bibinfo {collaboration} {LISA}),\ }\bibfield  {title} {\bibinfo {title} {{Laser Interferometer Space Antenna}},\ }\href@noop {} {\  (\bibinfo {year} {2017})},\ \Eprint {https://arxiv.org/abs/1702.00786} {arXiv:1702.00786 [astro-ph.IM]} \BibitemShut {NoStop}%
\bibitem [{\citenamefont {Seto}\ \emph {et~al.}(2001)\citenamefont {Seto}, \citenamefont {Kawamura},\ and\ \citenamefont {Nakamura}}]{Seto:2001qf}%
  \BibitemOpen
  \bibfield  {author} {\bibinfo {author} {\bibfnamefont {N.}~\bibnamefont {Seto}}, \bibinfo {author} {\bibfnamefont {S.}~\bibnamefont {Kawamura}},\ and\ \bibinfo {author} {\bibfnamefont {T.}~\bibnamefont {Nakamura}},\ }\bibfield  {title} {\bibinfo {title} {{Possibility of direct measurement of the acceleration of the universe using 0.1-Hz band laser interferometer gravitational wave antenna in space}},\ }\href {https://doi.org/10.1103/PhysRevLett.87.221103} {\bibfield  {journal} {\bibinfo  {journal} {Phys. Rev. Lett.}\ }\textbf {\bibinfo {volume} {87}},\ \bibinfo {pages} {221103} (\bibinfo {year} {2001})},\ \Eprint {https://arxiv.org/abs/astro-ph/0108011} {arXiv:astro-ph/0108011} \BibitemShut {NoStop}%
\bibitem [{\citenamefont {Harry}\ \emph {et~al.}(2006)\citenamefont {Harry}, \citenamefont {Fritschel}, \citenamefont {Shaddock}, \citenamefont {Folkner},\ and\ \citenamefont {Phinney}}]{Harry:2006fi}%
  \BibitemOpen
  \bibfield  {author} {\bibinfo {author} {\bibfnamefont {G.~M.}\ \bibnamefont {Harry}}, \bibinfo {author} {\bibfnamefont {P.}~\bibnamefont {Fritschel}}, \bibinfo {author} {\bibfnamefont {D.~A.}\ \bibnamefont {Shaddock}}, \bibinfo {author} {\bibfnamefont {W.}~\bibnamefont {Folkner}},\ and\ \bibinfo {author} {\bibfnamefont {E.~S.}\ \bibnamefont {Phinney}},\ }\bibfield  {title} {\bibinfo {title} {{Laser interferometry for the big bang observer}},\ }\href {https://doi.org/10.1088/0264-9381/23/15/008} {\bibfield  {journal} {\bibinfo  {journal} {Class. Quant. Grav.}\ }\textbf {\bibinfo {volume} {23}},\ \bibinfo {pages} {4887} (\bibinfo {year} {2006})},\ \bibinfo {note} {[Erratum: Class.Quant.Grav. 23, 7361 (2006)]}\BibitemShut {NoStop}%
\bibitem [{\citenamefont {Aggarwal}\ \emph {et~al.}(2021)\citenamefont {Aggarwal} \emph {et~al.}}]{Aggarwal:2020olq}%
  \BibitemOpen
  \bibfield  {author} {\bibinfo {author} {\bibfnamefont {N.}~\bibnamefont {Aggarwal}} \emph {et~al.},\ }\bibfield  {title} {\bibinfo {title} {{Challenges and opportunities of gravitational-wave searches at MHz to GHz frequencies}},\ }\href {https://doi.org/10.1007/s41114-021-00032-5} {\bibfield  {journal} {\bibinfo  {journal} {Living Rev. Rel.}\ }\textbf {\bibinfo {volume} {24}},\ \bibinfo {pages} {4} (\bibinfo {year} {2021})},\ \Eprint {https://arxiv.org/abs/2011.12414} {arXiv:2011.12414 [gr-qc]} \BibitemShut {NoStop}%
\bibitem [{\citenamefont {Herman}\ \emph {et~al.}(2021)\citenamefont {Herman}, \citenamefont {F\"uzfa}, \citenamefont {Lehoucq},\ and\ \citenamefont {Clesse}}]{Herman:2020wao}%
  \BibitemOpen
  \bibfield  {author} {\bibinfo {author} {\bibfnamefont {N.}~\bibnamefont {Herman}}, \bibinfo {author} {\bibfnamefont {A.}~\bibnamefont {F\"uzfa}}, \bibinfo {author} {\bibfnamefont {L.}~\bibnamefont {Lehoucq}},\ and\ \bibinfo {author} {\bibfnamefont {S.}~\bibnamefont {Clesse}},\ }\bibfield  {title} {\bibinfo {title} {{Detecting planetary-mass primordial black holes with resonant electromagnetic gravitational-wave detectors}},\ }\href {https://doi.org/10.1103/PhysRevD.104.023524} {\bibfield  {journal} {\bibinfo  {journal} {Phys. Rev. D}\ }\textbf {\bibinfo {volume} {104}},\ \bibinfo {pages} {023524} (\bibinfo {year} {2021})},\ \Eprint {https://arxiv.org/abs/2012.12189} {arXiv:2012.12189 [gr-qc]} \BibitemShut {NoStop}%
\bibitem [{\citenamefont {Berlin}\ \emph {et~al.}(2022)\citenamefont {Berlin}, \citenamefont {Blas}, \citenamefont {Tito~D'Agnolo}, \citenamefont {Ellis}, \citenamefont {Harnik}, \citenamefont {Kahn},\ and\ \citenamefont {Sch\"utte-Engel}}]{Berlin:2021txa}%
  \BibitemOpen
  \bibfield  {author} {\bibinfo {author} {\bibfnamefont {A.}~\bibnamefont {Berlin}}, \bibinfo {author} {\bibfnamefont {D.}~\bibnamefont {Blas}}, \bibinfo {author} {\bibfnamefont {R.}~\bibnamefont {Tito~D'Agnolo}}, \bibinfo {author} {\bibfnamefont {S.~A.~R.}\ \bibnamefont {Ellis}}, \bibinfo {author} {\bibfnamefont {R.}~\bibnamefont {Harnik}}, \bibinfo {author} {\bibfnamefont {Y.}~\bibnamefont {Kahn}},\ and\ \bibinfo {author} {\bibfnamefont {J.}~\bibnamefont {Sch\"utte-Engel}},\ }\bibfield  {title} {\bibinfo {title} {{Detecting high-frequency gravitational waves with microwave cavities}},\ }\href {https://doi.org/10.1103/PhysRevD.105.116011} {\bibfield  {journal} {\bibinfo  {journal} {Phys. Rev. D}\ }\textbf {\bibinfo {volume} {105}},\ \bibinfo {pages} {116011} (\bibinfo {year} {2022})},\ \Eprint {https://arxiv.org/abs/2112.11465} {arXiv:2112.11465 [hep-ph]} \BibitemShut {NoStop}%
\bibitem [{\citenamefont {Domcke}\ \emph {et~al.}(2022)\citenamefont {Domcke}, \citenamefont {Garcia-Cely},\ and\ \citenamefont {Rodd}}]{Domcke:2022rgu}%
  \BibitemOpen
  \bibfield  {author} {\bibinfo {author} {\bibfnamefont {V.}~\bibnamefont {Domcke}}, \bibinfo {author} {\bibfnamefont {C.}~\bibnamefont {Garcia-Cely}},\ and\ \bibinfo {author} {\bibfnamefont {N.~L.}\ \bibnamefont {Rodd}},\ }\bibfield  {title} {\bibinfo {title} {{Novel Search for High-Frequency Gravitational Waves with Low-Mass Axion Haloscopes}},\ }\href {https://doi.org/10.1103/PhysRevLett.129.041101} {\bibfield  {journal} {\bibinfo  {journal} {Phys. Rev. Lett.}\ }\textbf {\bibinfo {volume} {129}},\ \bibinfo {pages} {041101} (\bibinfo {year} {2022})},\ \Eprint {https://arxiv.org/abs/2202.00695} {arXiv:2202.00695 [hep-ph]} \BibitemShut {NoStop}%
\bibitem [{\citenamefont {Herman}\ \emph {et~al.}()\citenamefont {Herman}, \citenamefont {Lehoucq},\ and\ \citenamefont {F\'{u}zfa}}]{Herman2022}%
  \BibitemOpen
  \bibfield  {author} {\bibinfo {author} {\bibfnamefont {N.}~\bibnamefont {Herman}}, \bibinfo {author} {\bibfnamefont {L.}~\bibnamefont {Lehoucq}},\ and\ \bibinfo {author} {\bibfnamefont {A.}~\bibnamefont {F\'{u}zfa}},\ }\bibfield  {title} {\bibinfo {title} {{Electromagnetic Antennas for the Resonant Detection of the Stochastic Gravitational Wave Background}},\ }\href@noop {} {\ }\Eprint {https://arxiv.org/abs/2203.15668} {arXiv:2203.15668 [gr-qc]} \BibitemShut {NoStop}%
\bibitem [{\citenamefont {Bringmann}\ \emph {et~al.}(2023)\citenamefont {Bringmann}, \citenamefont {Domcke}, \citenamefont {Fuchs},\ and\ \citenamefont {Kopp}}]{Bringmann:2023gba}%
  \BibitemOpen
  \bibfield  {author} {\bibinfo {author} {\bibfnamefont {T.}~\bibnamefont {Bringmann}}, \bibinfo {author} {\bibfnamefont {V.}~\bibnamefont {Domcke}}, \bibinfo {author} {\bibfnamefont {E.}~\bibnamefont {Fuchs}},\ and\ \bibinfo {author} {\bibfnamefont {J.}~\bibnamefont {Kopp}},\ }\bibfield  {title} {\bibinfo {title} {{High-frequency gravitational wave detection via optical frequency modulation}},\ }\href {https://doi.org/10.1103/PhysRevD.108.L061303} {\bibfield  {journal} {\bibinfo  {journal} {Phys. Rev. D}\ }\textbf {\bibinfo {volume} {108}},\ \bibinfo {pages} {L061303} (\bibinfo {year} {2023})},\ \Eprint {https://arxiv.org/abs/2304.10579} {arXiv:2304.10579 [hep-ph]} \BibitemShut {NoStop}%
\bibitem [{\citenamefont {Starobinsky}(1980)}]{Starobinsky1980}%
  \BibitemOpen
  \bibfield  {author} {\bibinfo {author} {\bibfnamefont {A.~A.}\ \bibnamefont {Starobinsky}},\ }\bibfield  {title} {\bibinfo {title} {{A New Type of Isotropic Cosmological Models Without Singularity}},\ }\href {https://doi.org/10.1016/0370-2693(80)90670-X} {\bibfield  {journal} {\bibinfo  {journal} {Phys. Lett. B}\ }\textbf {\bibinfo {volume} {91}},\ \bibinfo {pages} {99} (\bibinfo {year} {1980})}\BibitemShut {NoStop}%
\bibitem [{\citenamefont {Garcia}\ \emph {et~al.}(2020{\natexlab{a}})\citenamefont {Garcia}, \citenamefont {Kaneta}, \citenamefont {Mambrini},\ and\ \citenamefont {Olive}}]{Garcia2020}%
  \BibitemOpen
  \bibfield  {author} {\bibinfo {author} {\bibfnamefont {M.~A.~G.}\ \bibnamefont {Garcia}}, \bibinfo {author} {\bibfnamefont {K.}~\bibnamefont {Kaneta}}, \bibinfo {author} {\bibfnamefont {Y.}~\bibnamefont {Mambrini}},\ and\ \bibinfo {author} {\bibfnamefont {K.~A.}\ \bibnamefont {Olive}},\ }\bibfield  {title} {\bibinfo {title} {Reheating and post-inflationary production of dark matter},\ }\bibfield  {journal} {\bibinfo  {journal} {Phys. Rev. D 101, 123507 (2020)}\ }\href {https://doi.org/10.1103/PhysRevD.101.123507} {10.1103/PhysRevD.101.123507} (\bibinfo {year} {2020}{\natexlab{a}}),\ \Eprint {https://arxiv.org/abs/2004.08404} {arXiv:2004.08404 [hep-ph]} \BibitemShut {NoStop}%
\bibitem [{\citenamefont {Garcia}\ \emph {et~al.}(2020{\natexlab{b}})\citenamefont {Garcia}, \citenamefont {Kaneta}, \citenamefont {Mambrini},\ and\ \citenamefont {Olive}}]{Garcia2020a}%
  \BibitemOpen
  \bibfield  {author} {\bibinfo {author} {\bibfnamefont {M.~A.~G.}\ \bibnamefont {Garcia}}, \bibinfo {author} {\bibfnamefont {K.}~\bibnamefont {Kaneta}}, \bibinfo {author} {\bibfnamefont {Y.}~\bibnamefont {Mambrini}},\ and\ \bibinfo {author} {\bibfnamefont {K.~A.}\ \bibnamefont {Olive}},\ }\bibfield  {title} {\bibinfo {title} {Inflaton oscillations and post-inflationary reheating},\ }\href {https://doi.org/10.1088/1475-7516/2021/04/012} {\bibfield  {journal} {\bibinfo  {journal} {Journal of Cosmology and Astroparticle Physics}\ }\textbf {\bibinfo {volume} {2021}}\bibfield  {number} {\bibinfo  {number} { (04)},\ \bibinfo {pages} {012}},\ }\Eprint {https://arxiv.org/abs/2012.10756} {arXiv:2012.10756 [hep-ph]} \BibitemShut {NoStop}%
\bibitem [{\citenamefont {Ellis}\ \emph {et~al.}(2015)\citenamefont {Ellis}, \citenamefont {Garcia}, \citenamefont {Nanopoulos},\ and\ \citenamefont {Olive}}]{Ellis2015}%
  \BibitemOpen
  \bibfield  {author} {\bibinfo {author} {\bibfnamefont {J.}~\bibnamefont {Ellis}}, \bibinfo {author} {\bibfnamefont {M.~A.~G.}\ \bibnamefont {Garcia}}, \bibinfo {author} {\bibfnamefont {D.~V.}\ \bibnamefont {Nanopoulos}},\ and\ \bibinfo {author} {\bibfnamefont {K.~A.}\ \bibnamefont {Olive}},\ }\bibfield  {title} {\bibinfo {title} {{Calculations of Inflaton Decays and Reheating: with Applications to No-Scale Inflation Models}},\ }\href {https://doi.org/10.1088/1475-7516/2015/07/050} {\bibfield  {journal} {\bibinfo  {journal} {JCAP}\ }\textbf {\bibinfo {volume} {07}},\ \bibinfo {pages} {050}},\ \Eprint {https://arxiv.org/abs/1505.06986} {arXiv:1505.06986 [hep-ph]} \BibitemShut {NoStop}%
\bibitem [{\citenamefont {Garcia}\ and\ \citenamefont {Amin}(2018)}]{Garcia:2018wtq}%
  \BibitemOpen
  \bibfield  {author} {\bibinfo {author} {\bibfnamefont {M.~A.~G.}\ \bibnamefont {Garcia}}\ and\ \bibinfo {author} {\bibfnamefont {M.~A.}\ \bibnamefont {Amin}},\ }\bibfield  {title} {\bibinfo {title} {{Prethermalization production of dark matter}},\ }\href {https://doi.org/10.1103/PhysRevD.98.103504} {\bibfield  {journal} {\bibinfo  {journal} {Phys. Rev. D}\ }\textbf {\bibinfo {volume} {98}},\ \bibinfo {pages} {103504} (\bibinfo {year} {2018})},\ \Eprint {https://arxiv.org/abs/1806.01865} {arXiv:1806.01865 [hep-ph]} \BibitemShut {NoStop}%
\bibitem [{\citenamefont {Ballesteros}\ \emph {et~al.}(2021)\citenamefont {Ballesteros}, \citenamefont {Garcia},\ and\ \citenamefont {Pierre}}]{Ballesteros:2020adh}%
  \BibitemOpen
  \bibfield  {author} {\bibinfo {author} {\bibfnamefont {G.}~\bibnamefont {Ballesteros}}, \bibinfo {author} {\bibfnamefont {M.~A.~G.}\ \bibnamefont {Garcia}},\ and\ \bibinfo {author} {\bibfnamefont {M.}~\bibnamefont {Pierre}},\ }\bibfield  {title} {\bibinfo {title} {{How warm are non-thermal relics? Lyman-$\alpha$ bounds on out-of-equilibrium dark matter}},\ }\href {https://doi.org/10.1088/1475-7516/2021/03/101} {\bibfield  {journal} {\bibinfo  {journal} {JCAP}\ }\textbf {\bibinfo {volume} {03}},\ \bibinfo {pages} {101}},\ \Eprint {https://arxiv.org/abs/2011.13458} {arXiv:2011.13458 [hep-ph]} \BibitemShut {NoStop}%
\bibitem [{\citenamefont {Garcia}\ \emph {et~al.}(2022)\citenamefont {Garcia}, \citenamefont {Kaneta}, \citenamefont {Mambrini}, \citenamefont {Olive},\ and\ \citenamefont {Verner}}]{Garcia:2021iag}%
  \BibitemOpen
  \bibfield  {author} {\bibinfo {author} {\bibfnamefont {M.~A.~G.}\ \bibnamefont {Garcia}}, \bibinfo {author} {\bibfnamefont {K.}~\bibnamefont {Kaneta}}, \bibinfo {author} {\bibfnamefont {Y.}~\bibnamefont {Mambrini}}, \bibinfo {author} {\bibfnamefont {K.~A.}\ \bibnamefont {Olive}},\ and\ \bibinfo {author} {\bibfnamefont {S.}~\bibnamefont {Verner}},\ }\bibfield  {title} {\bibinfo {title} {{Freeze-in from preheating}},\ }\href {https://doi.org/10.1088/1475-7516/2022/03/016} {\bibfield  {journal} {\bibinfo  {journal} {JCAP}\ }\textbf {\bibinfo {volume} {03}}\bibfield  {number} {\bibinfo  {number} { (03)},\ \bibinfo {pages} {016}},\ }\Eprint {https://arxiv.org/abs/2109.13280} {arXiv:2109.13280 [hep-ph]} \BibitemShut {NoStop}%
\bibitem [{\citenamefont {Ema}\ \emph {et~al.}(2022)\citenamefont {Ema}, \citenamefont {Mukaida},\ and\ \citenamefont {Nakayama}}]{Ema:2021fdz}%
  \BibitemOpen
  \bibfield  {author} {\bibinfo {author} {\bibfnamefont {Y.}~\bibnamefont {Ema}}, \bibinfo {author} {\bibfnamefont {K.}~\bibnamefont {Mukaida}},\ and\ \bibinfo {author} {\bibfnamefont {K.}~\bibnamefont {Nakayama}},\ }\bibfield  {title} {\bibinfo {title} {{Scalar field couplings to quadratic curvature and decay into gravitons}},\ }\href {https://doi.org/10.1007/JHEP05(2022)087} {\bibfield  {journal} {\bibinfo  {journal} {JHEP}\ }\textbf {\bibinfo {volume} {05}},\ \bibinfo {pages} {087}},\ \Eprint {https://arxiv.org/abs/2112.12774} {arXiv:2112.12774 [hep-ph]} \BibitemShut {NoStop}%
\bibitem [{\citenamefont {Aghanim}\ \emph {et~al.}(2020)\citenamefont {Aghanim} \emph {et~al.}}]{Planck:2018vyg}%
  \BibitemOpen
  \bibfield  {author} {\bibinfo {author} {\bibfnamefont {N.}~\bibnamefont {Aghanim}} \emph {et~al.} (\bibinfo {collaboration} {Planck}),\ }\bibfield  {title} {\bibinfo {title} {{Planck 2018 results. VI. Cosmological parameters}},\ }\href {https://doi.org/10.1051/0004-6361/201833910} {\bibfield  {journal} {\bibinfo  {journal} {Astron. Astrophys.}\ }\textbf {\bibinfo {volume} {641}},\ \bibinfo {pages} {A6} (\bibinfo {year} {2020})},\ \bibinfo {note} {[Erratum: Astron.Astrophys. 652, C4 (2021)]},\ \Eprint {https://arxiv.org/abs/1807.06209} {arXiv:1807.06209 [astro-ph.CO]} \BibitemShut {NoStop}%
\bibitem [{\citenamefont {Abbott}\ \emph {et~al.}(2018)\citenamefont {Abbott} \emph {et~al.}}]{KAGRA:2013rdx}%
  \BibitemOpen
  \bibfield  {author} {\bibinfo {author} {\bibfnamefont {B.~P.}\ \bibnamefont {Abbott}} \emph {et~al.} (\bibinfo {collaboration} {KAGRA, LIGO Scientific, Virgo, VIRGO}),\ }\bibfield  {title} {\bibinfo {title} {{Prospects for observing and localizing gravitational-wave transients with Advanced LIGO, Advanced Virgo and KAGRA}},\ }\href {https://doi.org/10.1007/s41114-020-00026-9} {\bibfield  {journal} {\bibinfo  {journal} {Living Rev. Rel.}\ }\textbf {\bibinfo {volume} {21}},\ \bibinfo {pages} {3} (\bibinfo {year} {2018})},\ \Eprint {https://arxiv.org/abs/1304.0670} {arXiv:1304.0670 [gr-qc]} \BibitemShut {NoStop}%
\bibitem [{\citenamefont {Thrane}\ and\ \citenamefont {Romano}(2013)}]{Thrane:2013oya}%
  \BibitemOpen
  \bibfield  {author} {\bibinfo {author} {\bibfnamefont {E.}~\bibnamefont {Thrane}}\ and\ \bibinfo {author} {\bibfnamefont {J.~D.}\ \bibnamefont {Romano}},\ }\bibfield  {title} {\bibinfo {title} {{Sensitivity curves for searches for gravitational-wave backgrounds}},\ }\href {https://doi.org/10.1103/PhysRevD.88.124032} {\bibfield  {journal} {\bibinfo  {journal} {Phys. Rev. D}\ }\textbf {\bibinfo {volume} {88}},\ \bibinfo {pages} {124032} (\bibinfo {year} {2013})},\ \Eprint {https://arxiv.org/abs/1310.5300} {arXiv:1310.5300 [astro-ph.IM]} \BibitemShut {NoStop}%
\bibitem [{\citenamefont {Reitze}\ \emph {et~al.}(2019)\citenamefont {Reitze} \emph {et~al.}}]{Reitze:2019iox}%
  \BibitemOpen
  \bibfield  {author} {\bibinfo {author} {\bibfnamefont {D.}~\bibnamefont {Reitze}} \emph {et~al.},\ }\bibfield  {title} {\bibinfo {title} {{Cosmic Explorer: The U.S. Contribution to Gravitational-Wave Astronomy beyond LIGO}},\ }\href@noop {} {\bibfield  {journal} {\bibinfo  {journal} {Bull. Am. Astron. Soc.}\ }\textbf {\bibinfo {volume} {51}},\ \bibinfo {pages} {035} (\bibinfo {year} {2019})},\ \Eprint {https://arxiv.org/abs/1907.04833} {arXiv:1907.04833 [astro-ph.IM]} \BibitemShut {NoStop}%
\bibitem [{\citenamefont {Yeh}\ \emph {et~al.}(2022)\citenamefont {Yeh}, \citenamefont {Shelton}, \citenamefont {Olive},\ and\ \citenamefont {Fields}}]{Yeh:2022heq}%
  \BibitemOpen
  \bibfield  {author} {\bibinfo {author} {\bibfnamefont {T.-H.}\ \bibnamefont {Yeh}}, \bibinfo {author} {\bibfnamefont {J.}~\bibnamefont {Shelton}}, \bibinfo {author} {\bibfnamefont {K.~A.}\ \bibnamefont {Olive}},\ and\ \bibinfo {author} {\bibfnamefont {B.~D.}\ \bibnamefont {Fields}},\ }\bibfield  {title} {\bibinfo {title} {{Probing physics beyond the standard model: limits from BBN and the CMB independently and combined}},\ }\href {https://doi.org/10.1088/1475-7516/2022/10/046} {\bibfield  {journal} {\bibinfo  {journal} {JCAP}\ }\textbf {\bibinfo {volume} {10}},\ \bibinfo {pages} {046}},\ \Eprint {https://arxiv.org/abs/2207.13133} {arXiv:2207.13133 [astro-ph.CO]} \BibitemShut {NoStop}%
\bibitem [{\citenamefont {Hasegawa}\ \emph {et~al.}(2019)\citenamefont {Hasegawa}, \citenamefont {Hiroshima}, \citenamefont {Kohri}, \citenamefont {Hansen}, \citenamefont {Tram},\ and\ \citenamefont {Hannestad}}]{Hasegawa:2019jsa}%
  \BibitemOpen
  \bibfield  {author} {\bibinfo {author} {\bibfnamefont {T.}~\bibnamefont {Hasegawa}}, \bibinfo {author} {\bibfnamefont {N.}~\bibnamefont {Hiroshima}}, \bibinfo {author} {\bibfnamefont {K.}~\bibnamefont {Kohri}}, \bibinfo {author} {\bibfnamefont {R.~S.~L.}\ \bibnamefont {Hansen}}, \bibinfo {author} {\bibfnamefont {T.}~\bibnamefont {Tram}},\ and\ \bibinfo {author} {\bibfnamefont {S.}~\bibnamefont {Hannestad}},\ }\bibfield  {title} {\bibinfo {title} {{MeV-scale reheating temperature and thermalization of oscillating neutrinos by radiative and hadronic decays of massive particles}},\ }\href {https://doi.org/10.1088/1475-7516/2019/12/012} {\bibfield  {journal} {\bibinfo  {journal} {JCAP}\ }\textbf {\bibinfo {volume} {12}},\ \bibinfo {pages} {012}},\ \Eprint {https://arxiv.org/abs/1908.10189} {arXiv:1908.10189 [hep-ph]} \BibitemShut {NoStop}%
\bibitem [{\citenamefont {Pagano}\ \emph {et~al.}(2016)\citenamefont {Pagano}, \citenamefont {Salvati},\ and\ \citenamefont {Melchiorri}}]{Pagano:2015hma}%
  \BibitemOpen
  \bibfield  {author} {\bibinfo {author} {\bibfnamefont {L.}~\bibnamefont {Pagano}}, \bibinfo {author} {\bibfnamefont {L.}~\bibnamefont {Salvati}},\ and\ \bibinfo {author} {\bibfnamefont {A.}~\bibnamefont {Melchiorri}},\ }\bibfield  {title} {\bibinfo {title} {{New constraints on primordial gravitational waves from Planck 2015}},\ }\href {https://doi.org/10.1016/j.physletb.2016.07.078} {\bibfield  {journal} {\bibinfo  {journal} {Phys. Lett. B}\ }\textbf {\bibinfo {volume} {760}},\ \bibinfo {pages} {823} (\bibinfo {year} {2016})},\ \Eprint {https://arxiv.org/abs/1508.02393} {arXiv:1508.02393 [astro-ph.CO]} \BibitemShut {NoStop}%
\bibitem [{\citenamefont {Ibanez}\ and\ \citenamefont {Ross}(1982)}]{Ibanez1982}%
  \BibitemOpen
  \bibfield  {author} {\bibinfo {author} {\bibfnamefont {L.~E.}\ \bibnamefont {Ibanez}}\ and\ \bibinfo {author} {\bibfnamefont {G.~G.}\ \bibnamefont {Ross}},\ }\bibfield  {title} {\bibinfo {title} {{SU(2)-L x U(1) Symmetry Breaking as a Radiative Effect of Supersymmetry Breaking in Guts}},\ }\href {https://doi.org/10.1016/0370-2693(82)91239-4} {\bibfield  {journal} {\bibinfo  {journal} {Phys. Lett. B}\ }\textbf {\bibinfo {volume} {110}},\ \bibinfo {pages} {215} (\bibinfo {year} {1982})}\BibitemShut {NoStop}%
\bibitem [{\citenamefont {Inoue}\ \emph {et~al.}(1982{\natexlab{a}})\citenamefont {Inoue}, \citenamefont {Kakuto}, \citenamefont {Komatsu},\ and\ \citenamefont {Takeshita}}]{Inoue1982}%
  \BibitemOpen
  \bibfield  {author} {\bibinfo {author} {\bibfnamefont {K.}~\bibnamefont {Inoue}}, \bibinfo {author} {\bibfnamefont {A.}~\bibnamefont {Kakuto}}, \bibinfo {author} {\bibfnamefont {H.}~\bibnamefont {Komatsu}},\ and\ \bibinfo {author} {\bibfnamefont {S.}~\bibnamefont {Takeshita}},\ }\bibfield  {title} {\bibinfo {title} {{Low-Energy Parameters and Particle Masses in a Supersymmetric Grand Unified Model}},\ }\href {https://doi.org/10.1143/PTP.67.1889} {\bibfield  {journal} {\bibinfo  {journal} {Prog. Theor. Phys.}\ }\textbf {\bibinfo {volume} {67}},\ \bibinfo {pages} {1889} (\bibinfo {year} {1982}{\natexlab{a}})}\BibitemShut {NoStop}%
\bibitem [{\citenamefont {Inoue}\ \emph {et~al.}(1982{\natexlab{b}})\citenamefont {Inoue}, \citenamefont {Kakuto}, \citenamefont {Komatsu},\ and\ \citenamefont {Takeshita}}]{Inoue1982a}%
  \BibitemOpen
  \bibfield  {author} {\bibinfo {author} {\bibfnamefont {K.}~\bibnamefont {Inoue}}, \bibinfo {author} {\bibfnamefont {A.}~\bibnamefont {Kakuto}}, \bibinfo {author} {\bibfnamefont {H.}~\bibnamefont {Komatsu}},\ and\ \bibinfo {author} {\bibfnamefont {S.}~\bibnamefont {Takeshita}},\ }\bibfield  {title} {\bibinfo {title} {{Aspects of Grand Unified Models with Softly Broken Supersymmetry}},\ }\href {https://doi.org/10.1143/PTP.68.927} {\bibfield  {journal} {\bibinfo  {journal} {Prog. Theor. Phys.}\ }\textbf {\bibinfo {volume} {68}},\ \bibinfo {pages} {927} (\bibinfo {year} {1982}{\natexlab{b}})},\ \bibinfo {note} {[Erratum: Prog.Theor.Phys. 70, 330 (1983)]}\BibitemShut {NoStop}%
\bibitem [{\citenamefont {Alvarez-Gaume}\ \emph {et~al.}(1983)\citenamefont {Alvarez-Gaume}, \citenamefont {Polchinski},\ and\ \citenamefont {Wise}}]{AlvarezGaume1983}%
  \BibitemOpen
  \bibfield  {author} {\bibinfo {author} {\bibfnamefont {L.}~\bibnamefont {Alvarez-Gaume}}, \bibinfo {author} {\bibfnamefont {J.}~\bibnamefont {Polchinski}},\ and\ \bibinfo {author} {\bibfnamefont {M.~B.}\ \bibnamefont {Wise}},\ }\bibfield  {title} {\bibinfo {title} {{Minimal Low-Energy Supergravity}},\ }\href {https://doi.org/10.1016/0550-3213(83)90591-6} {\bibfield  {journal} {\bibinfo  {journal} {Nucl. Phys. B}\ }\textbf {\bibinfo {volume} {221}},\ \bibinfo {pages} {495} (\bibinfo {year} {1983})}\BibitemShut {NoStop}%
\bibitem [{\citenamefont {Ellis}\ \emph {et~al.}(2018)\citenamefont {Ellis}, \citenamefont {Gherghetta}, \citenamefont {Kaneta},\ and\ \citenamefont {Olive}}]{Ellis2018a}%
  \BibitemOpen
  \bibfield  {author} {\bibinfo {author} {\bibfnamefont {S.~A.~R.}\ \bibnamefont {Ellis}}, \bibinfo {author} {\bibfnamefont {T.}~\bibnamefont {Gherghetta}}, \bibinfo {author} {\bibfnamefont {K.}~\bibnamefont {Kaneta}},\ and\ \bibinfo {author} {\bibfnamefont {K.~A.}\ \bibnamefont {Olive}},\ }\bibfield  {title} {\bibinfo {title} {{New Weak-Scale Physics from SO(10) with High-Scale Supersymmetry}},\ }\href {https://doi.org/10.1103/PhysRevD.98.055009} {\bibfield  {journal} {\bibinfo  {journal} {Phys. Rev. D}\ }\textbf {\bibinfo {volume} {98}},\ \bibinfo {pages} {055009} (\bibinfo {year} {2018})},\ \Eprint {https://arxiv.org/abs/1807.06488} {arXiv:1807.06488 [hep-ph]} \BibitemShut {NoStop}%
\bibitem [{\citenamefont {Carney}\ \emph {et~al.}(2024)\citenamefont {Carney}, \citenamefont {Domcke},\ and\ \citenamefont {Rodd}}]{Carney:2023nzz}%
  \BibitemOpen
  \bibfield  {author} {\bibinfo {author} {\bibfnamefont {D.}~\bibnamefont {Carney}}, \bibinfo {author} {\bibfnamefont {V.}~\bibnamefont {Domcke}},\ and\ \bibinfo {author} {\bibfnamefont {N.~L.}\ \bibnamefont {Rodd}},\ }\bibfield  {title} {\bibinfo {title} {{Graviton detection and the quantization of gravity}},\ }\href {https://doi.org/10.1103/PhysRevD.109.044009} {\bibfield  {journal} {\bibinfo  {journal} {Phys. Rev. D}\ }\textbf {\bibinfo {volume} {109}},\ \bibinfo {pages} {044009} (\bibinfo {year} {2024})},\ \Eprint {https://arxiv.org/abs/2308.12988} {arXiv:2308.12988 [hep-th]} \BibitemShut {NoStop}%
\bibitem [{\citenamefont {Ito}\ \emph {et~al.}(2024)\citenamefont {Ito}, \citenamefont {Kohri},\ and\ \citenamefont {Nakayama}}]{Ito:2023nkq}%
  \BibitemOpen
  \bibfield  {author} {\bibinfo {author} {\bibfnamefont {A.}~\bibnamefont {Ito}}, \bibinfo {author} {\bibfnamefont {K.}~\bibnamefont {Kohri}},\ and\ \bibinfo {author} {\bibfnamefont {K.}~\bibnamefont {Nakayama}},\ }\bibfield  {title} {\bibinfo {title} {{Gravitational Wave Search through Electromagnetic Telescopes}},\ }\href {https://doi.org/10.1093/ptep/ptae004} {\bibfield  {journal} {\bibinfo  {journal} {PTEP}\ }\textbf {\bibinfo {volume} {2024}},\ \bibinfo {pages} {023E03} (\bibinfo {year} {2024})},\ \Eprint {https://arxiv.org/abs/2309.14765} {arXiv:2309.14765 [gr-qc]} \BibitemShut {NoStop}%
\end{thebibliography}%

\end{document}